\documentclass[12pt]{article}

\usepackage{amsmath}

\newcommand{\bea}{\begin{eqnarray}}
\newcommand{\eea}{\end{eqnarray}}
\def \rt {{\rm t}}
\def \ci{\cite}

\newcommand{\p}[1]{(\ref{#1})}
\newcommand{\bt}[1]{{\bar t}}

\newcommand{\pbpi}[1]{\bar\psi_{#1}{\psi}_{#1}}

\def\pbp{{\bar\psi}\psi }
\def\bp{{\bar\psi}}
\def\etab{{\bar\eta}}
\def\thetab{{\bar\theta}}
\def\psib{{\bar\psi}}
\def\bpsi{{\bar\psi}}

\newcommand{\commentout}[1]{}

   \usepackage{amscd}
   \usepackage{amsfonts}
   \usepackage{amssymb}
   \usepackage{amsthm}
   \usepackage{epsfig}

\def\be{\begin{equation}}
\def\ee{\end{equation}}
\def\ba{\begin{eqnarray}}
\def\ea{\end{eqnarray}}

\def\a{\alpha}
\def\b{\beta}

\def\p{\partial}

\def\N{{\bf N}}

\def\P{{\bf P}}

\def\Tr{{\rm  Tr}}
\def\tr{{\rm  tr}}
\def\STr{{\rm  STr}}
\def\Str{{\rm  STr}}

\def\ket[#1]{\left|#1\right>}
\def\bra[#1]{\left<#1\right|}

\def \ww {{\rm w}}

\numberwithin{equation}{section} 
\parskip=\medskipamount

\input epsf
\usepackage{epsfig,color,latexsym}
\def \del{\partial}
\def \a {\alpha}

\def\g{\gamma}
\def\s{\sigma}
\def\z{\zeta}

\def\ov{\over}
\def\la{\label}

\def \om {\omega}
\def \cL {{\cal L}}

\def\b{\beta}
\def\l{\lambda}

\def \adss{$AdS_5 \times S^5$\ }

\def \ov {\over}
\def \s{\sigma}

\def \ha {{1 \over 2}}

\def \la{\label}

\def \k {\kappa}
\def\foot{\footnote}

\def \const {{\rm const}}

\def \L {\Lambda}

\def \rX {{\rm X}}

\def \LL {Landau-Lifshitz\ } 

\newcommand{\rf}[1]{(\ref{#1})}
\renewcommand{\theequation}{\thesection.\arabic{equation}}
\renewcommand{\thefootnote}{\fnsymbol{footnote}}

\def\appendix#1{
  \addtocounter{section}{1}
  \setcounter{equation}{0}
  \renewcommand{\thesection}{\Alph{section}}
  \section*{Appendix \thesection\protect\indent \parbox[t]{11.15cm}
  {#1} }
  \addcontentsline{toc}{section}{Appendix \thesection\ \ \ #1}
  }

\def \ci {\cite}

\def \foot {\footnote}
\def \bi{\bibitem}
\def \tr {{\rm tr}}
\def \ha {{1 \over 2}}
\def \td {\tilde}

\def \ci{\cite}
\def \N {{\cal N}}
\def \const {{\rm const}}

\def \n {\nu} 
\def \tl {{\tilde \lambda}}

\def \inti {{\int^{2\pi}_0 {d \sigma \ov 2 \pi}}}

\def \be {\bea}
\def \ee {\eea}

\def \Tr {{\rm Tr}}
\def \P {\Phi}

\def \m {\mu }

\def \bV  {{\bf V}}
\def \bC  {{\bf C}}\def \bD  {{\bf D}}
\def \cL   {{\cal  L}}
\def \psib {{\bar\psi}}
\def \STr {{\rm Str}} 
\def \Str {{\rm Str}} 
\def \cD {{\cal D}} 
\def \X {{\rm X}} 
\def \etab {\bar \eta}
\def \psib {\bar \psi}
\def \vn {\vec n}

\def \psib {{\bar \psi}}
\def \tl {{\tilde \lambda}}
\def \etap {{\eta_\perp}}

\def \z {\zeta} \def \bz {\bar \z} 
\def \bp {\bar \psi} 
\def \pb {\bar \psi} 
\def \bzz {\bar \z \z} 
\def \zb {\bar \zeta}

\begin{document}
\thispagestyle{empty}
\def\thefootnote{\fnsymbol{footnote}}\begin{flushright} 
hep-th/0503185\\ 
Imperial/TP/3-04/nn
\end{flushright}\vskip 0.5cm\begin{center}
\LARGE{\bf 
Super spin chain coherent state actions and  
$AdS_5 \times S^5$ superstring}
\end{center}\vskip 0.8cm\begin{center}{\large B. Stefa\'nski, jr.$^{1,}
$\footnote{E-mail address: {\tt b.stefanski@ic.ac.uk}}
and A.A. Tseytlin$^{2,1,}$\footnote{Also at Lebedev Institute, Moscow. }
}\vskip 0.2cm{\it $^1$ Theoretical Physics Group,  Blackett  Laboratory, \\
Imperial College,\\ London SW7 2BZ, U.K.}
\vskip 0.2cm
{\it $^2$  Physics Department, The Ohio State University, \\
Columbus, OH 43210-1106, USA}
\end{center}
\begin{abstract}\noindent

We consider a generalization of  
 leading-order matching of coherent state 
actions for semiclassical states on the super Yang-Mills
  and superstring  sides of the AdS/CFT
 duality to sectors with fermions. 
 In particular, we discuss 
the $SU(1|1)$ and $SU(2|3)$ sectors containing states 
with angular momentum $J$ in $S^5$ and spin in $AdS_5$.
 On the SYM  side, we start with the dilatation operator
  in the $SU(2|3)$ sector
having   super spin chain Hamiltonian interpretation 
and derive the corresponding coherent state action
which is   quartic in fermions. 
This action  has essentially the same 
``Landau-Lifshitz'' form as the action 
in the bosonic $SU(3)$ sector with  the 
target space $CP^2$ 
     replaced by 
  the projective superspace $CP^{2|2}$.
  We also discuss the complete $PSU(2,2|4)$ one-loop SYM
  spin chain  coherent state sigma model action. 
We then attempt to relate it to  the corresponding 
 truncation of the 
full $AdS_5 \times S^5$  
superstring action written  in a  light-cone gauge
where it   has  simple  quartic fermionic structure.  
 In particular, we find that  part 
of the superstring action describing $SU(1|1)$ 
sector reduces  
to an  action of a  massive  2d relativistic 
fermion, with the 
expansion in the effective coupling $\td
 \l = {\l \ov J^2}$  being equivalent
to a non-relativistic expansion.

\end{abstract}

\vfill
\setcounter{footnote}{0}

\def\thefootnote{\arabic{footnote}}

\newpage
\renewcommand{\theequation}{\thesection.\arabic{equation}}
\section{Introduction}

Recent progress in understanding the AdS/CFT duality beyond 
the supergravity sector
was inspired by the suggestion to consider a subsector of 
semiclassical string states (and near-by fluctuations)  that carry  large 
 quantum numbers \ci{bmn,gkp,ft1,ft2}  and by the 
 relation between the ${\cal N}=4$ super Yang-Mills dilatation operator 
 and integrable spin chains \ci{mz1,bks,bs}
 (for reviews and further references see, e.g.,
  \ci{ts1,ts2,beis}).

Here we shall concentrate  on a particular  approach 
(suggested in \ci{kru}
and developed in \ci{krt,hl1,st,mik12,kt,ital,mik3,hl2,kr2})
to comparing 
gauge theory (spin chain)  and string  theory  sides
 of the duality. It  
 is based on considering 
 a  low-energy effective action for coherent states 
of the ferromagnetic spin chain and relating it 
to a ``fast-motion'' limit 
of the string action.
In addition to explaining how a limit of
 string action ``emerges'' 
from the gauge theory dilatation operator, 
  this approach  also clarifies the  
identification of states on the two sides 
of the duality~\ci{kt,kr2} as well as
matching the integrable structures. 
For example, in the $SU(3)$ sector  containing 
states corresponding 
to   operators tr$( \Phi^{J_1}_1 \Phi^{J_2}_2\Phi^{J_3}_3)$ built out of
3 chiral combinations of  ${\cal N}=4$ SYM scalars
one finds from  the 1-loop spin chain Hamiltonian \ci{mz1} 
 the ``Landau-Lifshitz'' type 
action for coherent states defined on $CP^2$, and an equivalent action 
comes out (in the large  $J$ limit) of  
the bosonic part 
of the classical superstring action \ci{hl1,st}.

The motivation behind the present work  is 
 to try  to generalize   
previous discussions
of the matching of coherent state actions 
in  bosonic sectors  to sectors with fermions.
Previous interesting work in this  direction 
appeared in \ci{mik3,hl2}
where  quadratic order in fermions was considered. 
The full  one-loop  SYM   dilatation operator \ci{b1}
is a Hamiltonian of the  $PSU(2,2|4)$ super spin chain \ci{bs}, 
and it is of obvious importance  to understand in general 
a relation between the corresponding coherent state action and 
the \adss superstring  action   
\ci{MT} based on the  $PSU(2,2|4)/[SO(1,4) \times SO(5)]$ 
supercoset.
That may  help to further clarify the structure of 
superstring theory in this background and its connection to SYM theory.  

Let us note that, in  general, one should be
 comparing quantum gauge theory states 
on $R \times S^3$ to quantum string theory
 states in (global) $AdS_5\times S^5$.
In the ``semiclassical'' 
limit of large quantum numbers 
 it is  natural  to consider  coherent 
 states on both sides of the duality. 
 In the presence of fermions  one cannot 
 follow the bosonic pattern and directly compare 
 classical string solutions  
  to spin chain configurations:
 the classical solutions will dependent on Grassmann 
 parameters  and their 
  energy and charges will be even elements 
 of a Grassmann algebra.\foot{Related issues 
 were recently discussed in \ci{bksz,aat}.}
  To give an  interpretation to such 
 solutions  
  one would  need to 
 assign some expectation values to the Grassmann 
 elements so that they    approximate the results found 
 for the corresponding quantum states
 with some fermionic occupation numbers. 
 
 In order to  by-pass this complication one 
 may compare   not the states/solutions   but   
 semiclassical  effective actions with fermions 
 that appear in the relevant limits 
 both on the  spin chain side and the string theory side. 
 Indeed, one may 
  reformulate the spin-chain dynamics in terms of 
 a coherent-state path integral and then compare 
 the fermionic action  that appears there in a 
 continuum limit
 (and describing a particular class of 
 ``semiclassical'' states) 
 to a limit of superstring action  appearing in the string path
 integral. 
 Such a comparison of fermionic effective actions 
 is what we will be aiming at below, 
 but we will also discuss some of their Grassmann-valued 
 classical solutions.
 
On the  SYM or spin chain side,  we shall concentrate on the closed 
$SU(2|3)$ sector \ci{b2}  which generalizes the scalar  $SU(3)$ 
sector to include in the single-trace operators powers of 
two ``gluino'' fermionic components. We 
shall  systematically derive the corresponding (1-loop) 
coherent state action which has a natural 
 interpretation
as   a Landau-Lifshitz sigma model on the projective 
superspace 
$CP^{2|2}$ (an equivalent action was found independently 
in~\ci{hl2}).\footnote{See also a related
discussion in the very recent paper which studies
 the $SU(1,1|1)$ sector~\cite{bcm} and its relation to superstring action to quadratic order in 
fermions.}
On the string theory side, 
we shall  start with an  
explicit form of the  \adss action   
in  the    light-cone  $\kappa$-symmetry gauge 
of   \ci{mt,mtt}. This action  
 is at most quartic in fermions and has
  manifest $SU(4)$ symmetry.
 We shall discuss  how to 
 truncate this  action  to the $SU(2|3)$ sector 
 by first  writing it  in the $SU(3) \times U(1)$ 
 invariant form and then  isolating the 
 singlet fermionic sector.
 Understanding the issue  of consistent truncation
 to the  $SU(2|3)$ sector and also attempting to 
  including quartic
 fermionic terms  are novel elements of  the present  work. 
 The precise matching of the quartic fermionic terms 
 appears to depend on a particular choice of 
 field redefinitions that we did not succeed in finding.

To motivate the required  truncation  of the superstring action 
let us recall the 
contents of the $SU(2|3)$ sector on the SYM side \ci{b2,staud}. 
Starting with  the  $\N=4$ SYM  theory 
written in terms of the $\N=1$  superfields we may consider the 
operator $O$=tr$( \Phi^{J_1}_1 \Phi^{J_2}_2\Phi^{J_3}_3 \psi_1 ^{K_1}
 \psi_2^{K_2} )$ built out of three chiral scalars  of the ``matter'' 
 supermultiplets and  two spinor components
 of the ``gaugino'' supermultiplet 
 ($W_\a= \psi_\a + ...$, \ $\a=1,2$).
 Then we will have the  $SU(3)\subset SO(6) $ R-symmetry 
  acting on the scalars (under which the fermions 
   $\psi_\a$ are singlets) 
     and the   $SU(2)\subset SU(2,2) $  symmetry acting on the fermions 
 (under which the scalars are singlets). 
 The latter symmetry is essentially the 
 Lorentz spin  symmetry,  and $\psi_1 $ may be thought of 
 as a ``spin-up'', and $\psi_2$ as a  ``spin-down'' state.
  The above operator $O$ has canonical dimension 
 $ \Delta_0 = J  + {3\ov 2}( K_1 + K_2), $ with $ J= J_1 + J_2 + J_3 $
 being the total R-charge. Then 
  $S= \ha (K_1-K_2)$ is  the Lorentz spin
and $L= J + K_1 + K_2$ is  the total number of fields
or the length of the corresponding 
spin chain.\foot{Beyond one loop the length can fluctuate 
as one can trade a scalar $SU(3)$ singlet $ \epsilon^{ijk} \P_i \P_j \P_k$ 
for the $SU(2)$ singlet  $ \epsilon^{\a\b} \psi_\alpha \psi_\beta$
which has the same canonical dimension \ci{b2}. 
 }
  One may also consider various subsectors of the $SU(2|3)$ sector, 
for example, $SU(1|3)$  (3 scalars and 1 fermion).
The  simplest  subsector (which is closed to all orders \ci{b2,beis})
 is  the 
$SU(1|1)$  subsector  
 containing the operators tr$( \Phi^J \psi^K)$
with $ \Delta_0= J + { 3\ov 2} K = L + S , \  L= J + K, \ 
S= \ha K $, with the   $K=0$ case being  the BPS vacuum. 
 
 The corresponding string states should thus have
   both the $S^5$ angular
momenta (carried by the bosonic coordinates)  and 
one component of the  $AdS_5$ spin
 (carried by the fermionic coordinates). Also, the 
 two non-zero fermionic coordinates
 should be singlets under the $SU(3)$ R-symmetry.
In particular,  the closed $SU(1|1)$ sector should be 
described by an ``extension'' of the BMN point-like BPS state 
(carrying $S^5$ 
momentum $J$) 
 by a single fermion. The associated  coherent state action
will then  involve only a single 
fermionic  variable. 

\bigskip
\bigskip

We start in  section  2  by   
deriving  the coherent-state action
corresponding to the 1-loop  SYM dilatation operator
 in the $SU(2|3)$ sector.
We emphasize its geometrical interpretation 
as a Landau-Lifshitz 
sigma model on the  projective  superspace $CP^{2|2}$ 
and show that there exists a field 
redefinition that makes the action 
  quartic in fermions. 
 We mention then  a particular fermionic 
 classical solution  which generalizes 
 a static  bosonic $SU(2)$ Landau-Lifshitz
  solution (corresponding 
   to a circular spinning string with two equal angular
  momenta \ci{ft2,art}).
  We also 
  discuss the generalization to 
  the full  $PSU(2,2|4)$ spin chain  coherent state
  sigma model action.
  
In section 3 we consider  
  the \adss superstring  action \ci{MT} 
   in the light-cone $\kappa$-symmetry gauge  \ci{mt,mtt} which 
contains two fermionic coordinates transforming 
in the fundamental representation of $SU(4)$. 
We choose an ansatz for the $AdS_5$ bosonic
  coordinates that describes 
   strings localised at the center of  $AdS_5$ in 
global coordinates, 
with the global time proportional to the world-sheet time. 
Then  we rewrite the fermionic part of the action 
in the   manifestly $SU(3) \times U(1)$ form that no longer 
involves gamma-matrices. 
That  facilitates the 
 truncation to the $SU(2|3)$ sector 
where only two $SU(3)$-singlet fermionic coordinates are 
present. 

In section 4 we present   some classical 
solutions of the superstring action that are fermionic generalizations
of the  bosonic spinning string solutions. 
 That  helps to clarify possible consistent truncations of the
 superstring  equations of motion.

In section 5 we consider 
 the  matching of the  Landau-Lifshitz spin chain
 action to the  ``fast-string'' limit 
  of the  string  action. 
  In particular, we 
  consider the $SU(1|1)$ subsector  and relate the resulting fermionic
  action to that of a free {\it relativistic} 2d  fermion.

In Appendix A we  summarize our notation
and give useful gamma-matrix relations. 
In Appendix B we  present the  
expressions for $SU(4)$  charges of the 
 string  action.


\section{From spin chains to  sigma models:
 $SU(2|3)$ sector}


In this section we shall find the continuum limit of the coherent
 state expectation value 
of the one-loop $\cal N$=4 SYM dilatation operator in the  $SU(m|n)$ 
sub-sector  of the full $SU(2,2|4)$ dilatation operator. 
In our  choice of the  fundamental representation of $SU(n|m)=SU(m|n)$ 
  $n=0,1,2$
 will be   the number  of chiral fermionic (grassmann) 
 fields  and $m=1,2,3$  -- the number 
 of chiral scalar bosonic fields in the corresponding SYM single-trace operators 
\ci{b2}.\foot{In this section we shall often 
  use the notation  $SU(3|2)$ instead 
of $SU(2|3)$  used in 
 \ci{b2}. While the notation  $SU(2|3)$ (with $SU(2)\times SU(3)$
 subgroups being the space-time spin acting on fermions and internal
  R-symmetry  acting on bosons)
  is natural for a subgroup 
  of the full symmetry supergroup 
 $PSU(2,2|4)$ (with $SU(2,2) \times SU(4)$  bosonic subgroup
 being the product of the space-time conformal   symmetry and 
 the internal R-symmetry),  the ``reverse'' notation 
 $SU(3|2)$  seems more natural in discussing  the coset superspaces
 we will be interested  below. 
 We will choose  the  fundamental representation of $SU(m|n)$ 
 to contain 
$m$ ``bosons'' and $n$ ``fermions''. The superalgebra  
 $SU(m|n)$ can be realised in fundamental representation as a 
 set  of $(m+n) \times (m+n)$ matrices
  $M= \left(\begin{matrix}B & F\\ F' & B'  \end{matrix}\right)$, where 
the even  $m \times m $ 
 matrix  $B$ and $n \times n$  matrix  $B'$
are hermitian, with $\Str M = \tr B - \tr B'=0$, 
and the odd  matrices  satisfy $F^\dagger = F'$.
} 

Our aim will be to determine the structure 
of the associated low-energy effective actions 
for the coherent state fields.
The cases of the purely bosonic ($n=0$) $SU(2)$ and $SU(3)$ 
 sectors were discussed previously 
in \ci{kru,krt}   and in \ci{hl1,st}.
Related supergroup-type sigma  models were considered,  e.g., in 
~\cite{othersugroup}. Our final  result  for
 the coherent state action in the 
$SU(2|3)$ subsector will be the same as   in  \ci{hl2} 
but we shall emphasize the simple 
geometrical structure of the action. 
We shall also mention some classical fermionic solutions 
and a  generalization to the $PSU(2,2|4)$ 
 case (see  also \ci{mik3}). 

\subsection{Coherent state expectation value 
of  $SU(2|3)$ dilatation operator}

The  starting point will be   the  one-loop dilatation operator 
 in  the $SU(2|3)$  sector  which can be put into the form of  a spin 
 chain Hamiltonian ~\cite{b2}
\be
D
=\frac{2\lambda}{(4 \pi)^2}
\sum_{l=1}^{L}\left(1- P_{l,l+1}\right)\,.    \label{gsu}
\ee
Here  $ P_{l,l+1}$ is the  graded permutation operator, which 
acts by permuting a fermion or  boson   assigned to a  site $l$ with a fermion or boson  
assigned to  a site $l+1$
 with an additional minus sign if both fields  are fermionic.
The key observation is that the 
 permutation operator in the $SU(m|n)$ sector  can be 
expressed in terms of the  $SU(m|n)$ generators as~\footnote{When proving this 
identity  it is important to recall the definition of the tensor 
product  on a super-vector space:  $X^A\otimes X^f(v_f\otimes v_B)=
-X^A(v_f)\otimes X^f(v_B)$, where $X^f$ (or $v_f$) is a fermionic operator 
(or  vector) and $X^A$ ($v_B$) is any  bosonic or fermionic operator 
(or vector).}
\be
P_{l,l+1}=\frac{1}{m-n}+  \sum_{A,B} g_{AB} X^A_lX^B_{l+1}\,,
\ee
where $g_{AB}$ is the Cartan metric on the Lie superalgebra, i.e. the inverse of 
\be
g^{AB}=\STr (X^A X^B)\,.
\ee
For example, for the $SU(2)$ case ($m=2, \ n=0$) 
 with $X^A$ being the Pauli matrices $g_{AB}= {1 \ov 2} \delta_{AB}$ and 
$P_{l,l+1} = {1 \ov 2} (I + \sigma_l \cdot \sigma_{l+1})$.

Then  the dilatation operator in the $SU(m|n)$ sector is given 
explicitly by
\be
D=\frac{2\lambda}{(4 \pi)^2}\sum_{l=1}^{L}\left(\frac{m-n-1}{m-n}-
\sum_{A,B}g_{AB} X^A_lX^B_{l+1}\right)\,.\label{gsunm}
\ee
Let us first  consider   the simplest non-trivial $SU(2|1)$  sector 
where
\be
D_{_{SU(2|1)}}=- 
\frac{2 \lambda}{(4 \pi)^2}\sum_{l=1}^{L} \sum_{A,B}g_{AB} X^A_lX^B_{l+1}
\,,\label{gsu12}
\ee
and derive the 
corresponding 
coherent state effective action 
  (generalisations to  other $SU(m|n)$ sectors will be   straightforward).
We may choose the generators of $SU(2|1)$ in 
the fundamental representation 
as
\ba
X^1=\left(\begin{matrix}0&1&0 \\ 1&0&0 \\ 0&0&0 \end{matrix}\right), \ 
X^2=\left(\begin{matrix}0&-i&0 \\ i&0&0 \\ 0&0&0 \end{matrix}\right), \ 
X^3=\left(\begin{matrix}1&0&0 \\ 0&-1&0 \\ 0&0&0 \end{matrix}\right), \ 
X^4=\left(\begin{matrix}1&0&0 \\ 0&1&0 \\ 0&0&2 \end{matrix}\right), \nonumber\\
X^5=\left(\begin{matrix}0&0&1 \\ 0&0&0 \\ -1&0&0 \end{matrix}\right), \ 
X^6=\left(\begin{matrix}0&0&i \\ 0&0&0 \\ i&0&0 \end{matrix}\right), \ 
X^7=\left(\begin{matrix}0&0&0 \\ 0&0&1 \\ 0&-1&0 \end{matrix}\right), \ 
X^8=\left(\begin{matrix}0&0&0 \\ 0&0&i \\ 0&i&0 \end{matrix}\right)\,,\nonumber
\ea
 where $X^{1,2,3,4,8}$ are even and   $X^{5,6,7}$ are odd.\foot{We 
  take the even (b) and odd (f) 
generators  to satisfy
$
X_{(b)}^\dagger=X_{(b)}\,,\  \  X_{(f)}^\dagger=-X_{(f)}\,.
$} 
$X^3$ and $X^4$ form  Cartan subalgebra, and $X^1$, $X^2$ and $X^3$ 
form  the bosonic $SU(2)$ subgroup. 
As usual,   we 
can define a 
 coherent state by a ``rotation'' of a ``vacuum state'' 
by  the generators
 that do not preserve it ~\cite{sucoh} 
\be
\ket[N]\equiv {\cal N}e^{i(a_1 X^1+
a_2 X^2+\theta_1X^5+\theta_2X^6)}\ket[0]\,.\label{su12coh}
\ee
Here  $a_1,a_2$  and $\theta_1,\theta_2$ are real even 
and odd 
parameters 
and ${\cal N}$ is a (Grassmann-even) normalisation.\foot{Let us
recall 
that in the case of a free fermionic oscillator 
(or  Clifford  algebra) 
the fermionic coherent states  may be defined as 
$ \ket[\theta ] = e^{ -\theta a^\dagger} \ket[0], \ 
a \ket[0]=0, \ a a^\dagger + a^\dagger a =1, \ 
a \ket[\theta ] = \theta  \ket[\theta ].$}
We shall    choose the vacuum state to be  $\ket[0]=(1,0,0)$ 
which is  an eigen-state of 
the Cartan generators (and is annihilated by 
$X^7, X^8$)  -- it  corresponds to the BPS vacuum
 $\Tr(\Phi^L)$. 
 The coherent states are thus  parametrized by 
 the elements of the supercoset $G/H$ where $H$ is the stability subgroup of the
 vacuum, 
 i.e. by the points in the  projective superspace $CP^{1|1}
 =  SU(2|1)/[SU(1|1)\times U(1)]$.\footnote{Had we chosen
 instead   the ``fermionic'' vacuum  $(0,0,1)$
    (which corresponds to a non-BPS  state  $\Tr(\psi^L)$)
 we would get  the coset $SU(2|1)/[SU(2)\times U(1)]$.
}
Similarly,  we define 
\be
\bra[N]\equiv {\cal N}^*\bra[0]e^{-i(a X^1+bX^2-\theta_1X^5-\theta_2X^6)}\,.
\ee
In order to satisfy $
\left< N|N\right> =1$
we require
\be
{\cal N} =  {\cal N}^*=1-\frac{\sin 2\Delta}{\Delta}\theta^2 \,,
\ee
where
\be
\Delta\equiv \sqrt{a_1^2+a_2^2}\,,\qquad\theta^2\equiv\theta{\bar\theta}\,,\qquad \theta=\theta_1+i\theta_2\,,\qquad {\bar\theta}=\theta_1-i\theta_2\,,
\ee
\noindent Then   
\be
\bra[N]X^A\ket[N]=x^A \ , \ \ \ \ \ \ \ \ \ \ \ \ \ \
x^A= x^A(a_1,a_2,\theta_1,\theta_2) \ , 
\ee
 where  the explicit form of the functions $x^A$ is 
\ba
x^1 &=& \cos \varphi \left[-\sin2\Delta+\frac{\theta^2}{\Delta}
(\cos2\Delta+\frac{\sin2\Delta(4\sin^2\Delta-1)}{2\Delta})\right]\,,\nonumber \\ 
x^2&=&- \sin\varphi\left[-\sin2\Delta+ \frac{\theta^2}{\Delta}(\cos2\Delta+
\frac{\sin2\Delta(4\sin^2\Delta-1)}{2\Delta})\right]\,,\nonumber \\
x^3&=&\cos2\Delta+\frac{\theta^2}{\Delta}(\sin2\Delta-\frac{\sin\Delta\sin3\Delta}{\Delta})\,,\qquad
x^4=1+\frac{\theta^2}{\Delta^2}\sin^2\Delta\,,\nonumber \\
x^5&=&\frac{\sin 2\Delta}{2\Delta}\left({\bar\theta}-\theta\right)\,
,\qquad
x^6=i\frac{\sin 2\Delta}{2\Delta}\left({\bar\theta}+\theta\right)\,, \qquad
x^7=\frac{\sin^2\Delta}{\Delta}\left(\theta e^{-i\varphi}-{\bar\theta}e^{i\varphi}\right
)\,,\nonumber \\
x^8&=&-i\frac{\sin^2\Delta}{\Delta}\left
(\theta e^{-i\varphi}+{\bar\theta}e^{i\varphi}\right)\,,
\ \ \ \ \ \ \ \ \ \  e^{i\varphi}\equiv\frac{a_2+ia_1}{\Delta}\label{x1x8}
\ea
Let us  define the $SU(2|1)$  matrix $N$ 
belonging to the supercoset 
 $SU(2|1)/[SU(1|1) \times U(1) ] $     as
\be
N=  \sum_{A,B=1}^8 g_{AB}\ x^A  X^B\,, \ \ \ \ \ \ \  \ \ \ \ 
x^A=\STr(NX^A)  \ .   \label{Ndef}
\ee 
It is then easy to show that $N$ can be written as 
\be
N^q_{ \ p }  ={\bf V}^q {\bf V}_p-\delta_p^q \ ,\  \ \ \ \ \ \ \ \ 
 p,q=1,2,3  \ ,  
\label{nm}
\ee
where
\be
{\bf V}^p=(V_1,V_2,\psi)\,,\qquad
\bV_p \equiv ( \bV^p)^\dagger =(V_1^*,V_2^*,{\bar\psi})\,, \ \ \ \ \ \ 
 {\bf V}_p {\bf V}^p   = {\bf V}^p {\bf V}_p = 1\,, 
\ee
and thus 
\be  
N^\dagger = N \ , \ \ \ \ \  \ \ \ \  \STr N =0 \ , \ \ \ \ \ \ 
 \ \  N^2=- N  \ . 
\ee
It is assumed that the $_p ^{\ p}$ summation  (as in  $A_p B^p$)
 is done  with plus sign for the fermionic 
components, while the $^p _{\ p}$ summation (as in  $B^p A_p$) -- with 
minus  sign (so it is consistent  with the definition of the supertrace). 
The explicit form of  the Grassmann-valued constraint on $\bV^p$ is thus 
\be
|V_1|^2+|V_2|^2 +{\bar\psi}\psi  =1\,.\label{susph}
\ee
\noindent 
Both  $N$  and  ${\bf V}^p$ (the latter modulo $U(1)$  
phase transformations)
 thus  parametrise the supercoset $CP^{1|1}=SU(2|1)/[SU(1|1)\times U(1)]$.
The components of  ${\bf V}^p$  
can be expressed in terms of $\Delta$, $\varphi$ and $\theta$ in \rf{x1x8} as
\ba
V_1&=&-\cos\Delta-\frac{\theta^2}{2\Delta^2}\sin\Delta(\Delta-\sin2\Delta)\,,\\
V_2&=&e^{i\varphi}\big[\sin\Delta+\frac{\theta^2}{4\Delta^2}(\sin3\Delta-
\sin\Delta-2\Delta\cos\Delta)\big]\ , \\
\psi&=&\frac{\sin\Delta}{\Delta}\theta\,.
\ea
The above construction   
 is straightforward   to generalise to the  $SU(m|n)$ 
  case where the matrix $N$  should belong to 
   ($CP^{m-1}$ in the  bosonic  $n=0$ case \ci{st})
  $$CP^{m-1|n}=
 { SU(m|n)\ov SU(m-1|n )\times U(1)} $$ 
 i.e. 
\be 
N^q_{\ p } =(m-n) {\bf V}^q {\bf V}_p -\delta_p^q\  , \ \ \ \ \ \ \ \ \ \ 
  {\bf V}_p {\bf V}^p =1 \ , \la{his} \ee
\be N^\dagger =N\ , \ \ \ \  \STr N=0 \ , \ \ \ \ \ \ \ \ 
N^2=(m-n-2)N+(m-n-1)I\,.
\ee
The components of $\bV^p$ are  $V_i$ ($i=1,...,m$)
and $\psi_\a$ ($\a= 1,...,n$) 
with $\psib_\a\equiv  \psi^\dagger_\a$ 
\be  \bV^p= (V_i, \psi_\a) \ , \ \ \ \ \ 
  \ \ \ \ \ \ \ 
   V^*_i V_i +     \psib_\a \psi_\a=1 \ , \la{com}  \ee 
   where we assume summation over
repeated $i$ and $\a$ index.
   
Returning back to $SU(2|1)$ case let us  now  define     the coherent   state 
for the whole spin chain as 
\be
\ket[N]\equiv\prod_{l=1}^L\ket[N_l]\,,
\ee 
where $\ket[N_l]$ are  given by (\ref{su12coh}).
 Computing the matrix element of \rf{gsu12} we get 
\ba
\bra[N]D_{_{SU(2|1)}}\ket[N]&=&-   \frac{\lambda}{(4\pi)^2}\sum_{l=1}^L
g_{AB}\ \Str(N_lX^A)\  \Str(X^B N_{l+1})\nonumber \\
&=&\frac{\lambda}{(4\pi)^2}\sum_{l=1}^L
 \  \ha  g_{AB}\ \Str[(N_{l+1}-N_l)X^A] \  \Str[(N_{l+1}-N_l)X^B]\nonumber \\
&=&\frac{\lambda}{(4\pi)^2}\sum_{l=1}^L
\Str(N_{l+1}-N_l)^2 \ . 
\ea
We used the completeness relation
\be
\sum_{A,B}\ g_{AB}\ \STr(M X^A)\ \Str(X^B M )=2\ \STr M ^2\,,
\ee
valid for any matrix $M$ in the $SU(2|1)$ 
superalgebra and also that $\STr N^2_l= - \Str N_l =0$.  
Then  taking the relevant \ci{kru,krt,st} continuum limit
describing semiclassical low-energy states of the spin chain,
i.e.  $L\to \infty$  with 
 $ \tl\equiv {\lambda \ov L^2} 
$=fixed,   we get
\ba
\bra[N]D_{_{SU(2|1)}}\ket[N] \ 
&\rightarrow&
  L  \inti\    \frac{{\tilde\lambda}}{4}\ \STr(\p_1 N\p_1  N)\,. 
\ea
Rescaling $t\rightarrow \rt= \tl^{-1} t$,   the total
 coherent state  path integral action  becomes 
\be
I = L \int d\rt \inti \ {\cal L}_{_{SU(2|1)}} \ , \ \ \ \ \ \ \ \ \ 
{\cal L}_{_{SU(2|1)}}={\cal L}_{\mbox{\scriptsize WZ}}(N) 
-\frac{1}{4}\STr(\p_1N\p_1N)\,.
\label{lgsu12}
\ee
Here   ${\cal L}_{\mbox{\scriptsize WZ}}(N) $ is the usual WZ type term
(which can be computed as $ \bra[N] i \del_0 \ket[N]$)
\be
{\cal L}_{\mbox{\scriptsize WZ}}(N) =\frac{i}{2}\int_0^1dz\
 \Str\left(N [\p_zN,\p_0N\}\right)\,,
\ee
where $[\,\, \, ,\,\,\}$ is the superbracket.

\subsection{ $SU(2|3)$ Landau-Lifshitz  sigma model}

The Lagrangian in \rf{lgsu12}   admits a simpler local representation 
in terms of the vector variable $\bV^p$ with an additional $U(1)$ gauge symmetry
which 
is a direct generalization of the  $CP^{m-1}$   ``Landau-Lifshitz'' 
  Lagrangians in the bosonic $SU(2)$  and $SU(3)$ 
cases  in the form given in \ci{krt,st}: 
\be \la{bb}
\cL=-   iU^*_i\p_0 U_i   -\frac{1}{2}|D_1U_i|^2\ , \ \ \ \ \ \ \ \ \ \ \ \ \
|U_i |^2 =1 \ , \ \ \ \ee
\be   D_a U_i = \del_a U_i - i C_a U_i \ , \ \ \ \ \ \ 
C_a = -   iU^*_i\p_a U_i  \  , \ \ \ \ \ \ \ 
U^*_i D_a U_i =0 \ .  \la{uuu}  \ee
Here $U_i$ ($i=1,...,m$) belongs to $CP^{m-1} = SU(m)/[SU(m-1) \times U(1)]$:
in addition to the  unit modulus constraint the action  has gauge $U(1)$ symmetry. 
In the  supercoset case  we get the Lagrangian \rf{lgsu12}
defined on the  projective superspace (cf.
 \rf{his})\foot{Standard 2-d Lorentz-covariant sigma models 
   with projective superspaces as target spaces were 
   studied, e.g., in \ci{rs}.}
   $CP^{m-1|n} = SU(m|n)/[SU(m-1|n) \times U(1)] $
    (here $\bV_p= (\bV^p)^\dagger, \    \bD_1\bV_p =  (
   \bD_1\bV^p)^\dagger$)
 \be 
\cL= - i \bV_p \p_0 \bV^p   -\frac{1}{2} \bD_1\bV_p  \bD_1\bV^p
\ , \ \ \ \ \ \ \ \ \ \ \ 
 \bV_p   \bV^p=1 \ , \la{mai}  \ee
\be   \bD_a \bV^p = \del_a \bV^p  - i \bC_a \bV^p \ , \ \ \ \ \ \ \ \ 
\bC_a = -   i\bV_p \p_a \bV^p  \  .   \ee
Written explicitly in terms of the component fields in \rf{com}
this becomes 
\ba
{\cal L}&=&-iV^*_i\p_0 V_i-  i{\bar\psi_\a}\p_0\psi_\a 
-\frac{1}{2}\left(|{\cD}_1V_i|^2 +  {\cD}_1^*{\bar\psi_\a}{\cD}_1\psi_\a 
\right)\nonumber \\
&=&-iV^*_i\p_0V_i-  i{\bar\psi}_\a \p_0\psi_\a 
-\frac{1}{2}\left(|\p_1V_i|^2+\p_1{\bar\psi}_\a \p_1\psi_\a   -  \bC_1^2\right)
\,,\label{su12}
\ea
where
\be    \cD_a (V_i, \psi_\a ) = (\p_a - i \bC_{a}) ( V_i, \psi_\a) \ ,  
 \ \ \ \ \ \ \ \ \ \ \ \ \ \ \   \bC_{a}=
-iV_i^*\p_{a}V_i -    i{\bar\psi}_\a\p_{a}\psi_\a \,.
\la{conn}  \ee
It is convenient to decouple bosons and fermions in the constraint \rf{com}.
Let us first consider the $SU(2|1)$ sector and 
 define  the new bosonic field $U_i$ ($i=1,2$) by  
\be
V_i=U_i(1-\frac{1}{2}\psib {\psi})\,,\ \ \ \ \ \ \ \ \ \ \ \ \
U_i=V_i(1+\frac{1}{2}\psib {\psi})\,.\ 
\ee
Then the normalisation condition \rf{susph}  becomes simply 
\be
|U_1|^2+|U_2|^2=1\,, 
\ee
so that $U_i$  belongs to $CP^1$ (both $U_i$ and 
$\psi$ still transform under the $U(1)$ gauge symmetry).   
The gauge field in \rf{conn} is then
\be
\bC_{a}=C_{a}(1-\pbp)- \frac{i}{2}(\bp\p_{a}\psi + \psi   \p_{a}\bp )\,,\ \ \ \ \ \ \ 
C_{a}=-iU_i^*\p_{a}U_i\,.
\ee
and so the $SU(2|1)$ Lagrangian \rf{su12}  takes the form 
\ba
{\cal L}_{_{SU(2|1)}}&=&-iU^*_i\p_0U_i- i{\bar\psi}D_0\psi 
-\frac{1}{2}\left[(1-\pbp)|D_1U_i|^2+D_1^*{\bar\psi}D_1\psi  \right]\label{su21} \ , 
\ea
where $ D_{a}\equiv \p_{a}-iC_{a}$ is the ``bosonic'' covariant derivative. 
A remarkable feature of this Lagrangian is that there are no terms quartic
 in fermions:
 terms which come from $\bC_1^2$ and from $\p_1V_i\p_1V_i^*$ cancel.
This $CP^{1|1}$ supercoset Landau-Lifshitz   Lagrangian  is
 a generalization 
of the   $CP^1$  Lagrangian of the bosonic $SU(2)$ sector. 

Another   special case is the $CP^{0|1}$ model corresponding to 
the $SU(1|1)$ sector where we have only one component 
of $U_i$ ($|U_1|^2=1$)
which can  be gauge-fixed to  1. Then \rf{su21} 
 reduces to the abelian  \LL  system
\be\la{u11}
{\cal L}_{_{SU(1|1)}}=- i{\bp}\p_0\psi -  \frac{1}{2}\p_1{\bp}\p_1\psi\,.
\ee
We may repeat the same transformation 
 in  the $SU(3|2)$ case  \rf{su12} by solving  the 
 normalisation condition
   using the new bosonic fields $U_i$
\be
V_i=U_i\sqrt{1-\bp_\a{ \psi}_\a}=U_i\ \big[
1-\frac{1}{2}\bpsi_\a{\psi}_\a-\frac{1}{8}(\bpsi_\a{\psi}_\a)^2\big] \,,
\ \ \ \ \ \  \ \ \ \ \ \     |U_i|^2=1\  .    \label{newbos}
\ee
Then $\bC_a$ in \rf{conn} becomes 
\be
\bC_{a}=C_{a}\ (1-\pbpi\a)-  \frac{i}{2}(\bp_\a\p_{a}\psi_\a+\psi_\a\p_{a}\bp_\a)\,,
\ \ \ \ \ \ \ \ \ \ \ C_a =- iU_i^*\p_{a}U_i\, , 
\ee
and the Lagrangian \rf{su12} takes the form 
($i=1,2,3;\ \a=1,2$) 
\ba
{\cal L}_{_{SU(3|2)}}&=&-iU^*_i\p_0 U_i- i{\bar\psi}_\a D_0\psi_\a 
-\frac{1}{2}\big[ |D_1 U_i|^2(1-\pbpi\a)+D_1^*{\bar\psi}_\a D_1\psi_\a  
\nonumber \\
&&+\frac{1}{2}(\psi_\a D_1^*\bp_\a)^2+\frac{1}{2}(\bp_\a D_1\psi_\a)^2+\frac{1}{4}
\pbpi\a \p_1 (\pbpi\b)   \p_1 (\pbpi\g)     \big]\,. \la{six}
\ea
We may  recover the $SU(2|1)$ Lagrangian \rf{su21} 
by setting $U_3=0$ and $\psi_2=0$: the last sixth-order term 
(which originated from   $\p_1V_i\p_1V_i$)  then vanishes, and
 the quartic fermionic terms also  become    zero 
 since they are proportional to $\psi_\a\psi_\b$ or $\bp_\a\bp_\b$.


Since we would like  to compare the spin chain action 
 to the  superstring action which (in a  particular  gauge) contains 
 terms that are at most  quartic in 
fermions, it is useful to notice that one  can eliminate 
the sixth-order term by redefining the  fermionic field 
\be
\psi_\a \rightarrow \psi_\a  -\frac{1}{2}\ (\pbpi\b)       \psi_\a\, . 
\ee
Then 
\ba
{\cal L}_{_{SU(3|2)}}&=&-iU^*_i\p_0U_i- i(1- \pbpi\a  ){\bar\psi}_\a D_0\psi_\a \nonumber \\&&
-\frac{1}{2}\left[ (1- \pbpi\a + (\pbpi\a )^2 ) |D_1U _i|^2   
 + (1-  \pbpi\b  )D_1^*{\bar\psi}_\a  D_1\psi_\a 
\right.\nonumber \\&&\left.
+(\psi_\a D_1^*\bp_\a)(\bp_\b D_1\psi_\b)\right] \,.\label{su31}
\ea
In this form our $CP^{2|2}$ Landau-Lifshitz  Lagrangian  agrees  with the 
$SU(2|3)$ spin chain 
 coherent state Lagrangian found earlier in \ci{hl2}.

\subsection{An example of  fermionic solution
of the  Landau-Lifshitz model}\label{gaugesoln}

The bosonic $SU(2)$ \LL model has a very simple 
static solution \ci{krt}  which corresponds, in the string
 theory picture, 
 to the circular string rotating in $S^3$ part of $S^5$ with two equal
 angular momenta
  (i.e. X$_1 =\frac{1}{\sqrt{2}} e^{ i w \tau + i n \s},$ X$_2=
  \frac{1}{\sqrt{2}}e^{ i w \tau - i n \s}$):
   $U_{1} = \frac{1}{\sqrt{2}} e^{  i n \s},\  U_2=
  \frac{1}{\sqrt{2}}e^{ - i n \s}$.
  Interestingly, 
  the equations of motion that follow from the
   Lagrangian~(\ref{su21}) 
   ($\L$ is the Lagrange multiplier imposing $U^*_i U_i =1$)
\ba
0&=&-2i(1-  \pb  \psi)\p_0U_i+(1- \pb \psi )D_1^2U_i+\Lambda U_i\,,\\
0&=&-2iD_0\psi+|D_1U_i|^2\psi+D_1^2\psi\,.
\ea
admit the following generalization of the above 
bosonic static solution 
\ba
U_1&=&\frac{1}{\sqrt{2}}\left(e^{in\sigma}- e^{-in\sigma}\bz \z \right)\,,\\
U_2&=&\frac{1}{\sqrt{2}}\left(e^{-in\sigma}+ e^{in\sigma}\bz \z \right)\,,\\
\psi&=&e^{im\sigma}\z\,,
\ea
where $\z$ is a constant complex Grassmann parameter 
($U_i$ are even elements of the Grassmann algebra) 
 and $n\,,m$ are integers.
 Note that since our action has local 
 $U(1)$ symmetry (in addition to global $SU(2|1)$ symmetry)
 this solution is equivalent to the one with constant $\psi$
 field. 
 
For the above   ansatz
\be
C_1=-iU^*_i\p_1U_i=0\,,\qquad\mbox{and}\qquad|\p_1U_i|^2=n^2\,, 
\ee
and then 
\ba
0&=&(1- \bp \psi)\p_1^2U_i+\Lambda U_i\,,\\
0&=&n^2\psi+\p_1^2\psi\,.
\ea
The latter equation is solved if  $n=m$ 
while the former determines the Lagrange multiplier to be
\be
\Lambda=-(1-\bp \psi)U^*_i\p_1^2U_i=n^2(1- \bz \z )\,.
\ee
Since $\p_1^2U_i=-n^2U_i$, our ansatz does indeed satisfy the 
$U_i$ equation of motion. 

Surprisingly, the corresponding energy density (determined 
by the spatial derivative part of \rf{su21})
$ {\cal E} =
\frac{1}{2}\left[(1-\pbp)|D_1U_i|^2+D_1^*{\bar\psi}D_1\psi 
 \right]$,  evaluated on the 
 above solution,  is equal simply 
 to $ \ha n^2$, i.e. it 
 does not depend on $\zeta$.  Indeed, 
 one can 
 show that this solution can be obtained from the
 bosonic  $SU(2)$ subsector  solution
  by means of a global $SU(1|2)$
 rotation and a local $U(1)$ rotation.

\subsection{On  $PSU(2,2|4)$ \LL  model}

As was found  in \ci{st,mik12,mik3,kt}, 
 the generalization of the $CP^2$ 
\LL action   of the $SU(3)$ sector to the case of the $SO(6)$ 
sector is a similar action on the Grassmanian $G_{2,6}= SO(6)/[SO(4) \times SO(2)]$ 
which is the same as 
 a quadric in $CP^5$ defined by 
 $U_i U_i=0$ ($i=1,...,6$)  imposed  in addition to 
$U_i U_i^*=1$ and the $U(1)$ gauge invariance.
The discussion in \ci{mik3} suggests 
 that a generalization of the above $CP^{2|2}$
supercoset action \rf{mai} 
for the $SU(2|3)$  sector to the coherent state action 
for the  full $PSU(2,2|4)$ spin chain of \ci{bs} 
(with the vacuum chosen again  to 
represent  the BPS state $\tr \Phi^J$)  should be defined  on 
a 
super-Grassmanian which generalizes the product of the two bosonic Grassmanians 
$SO(2,4)/[SO(4) \times SO(2)]$ and  $ SO(6)/[SO(4) \times SO(2)]$ \ci{mik3}: 
$$G_{2|2, 4|4}=  SU(2,2|4)/[SU(2|2) \times SU(2|2)
 \times U(1) \times U(1) ]  \ . $$ 
The corresponding Lagrangian   is a direct 
generalization of   
\rf{mai}  where the analog of $\bV^p$ is
 subject to an additional $(\bV^p)^2=0$ constraint. To get 
an action with unconstrained fermions one would then 
 need only to redefine the bosons in a fashion
  similar to equation~(\ref{newbos}). 
  
It remains a challenge  to directly  relate this action 
to a limit of the superstring  action \ci{MT} defined on 
 the supercoset $SU(2,2|4)/[SO(1,4) 
\times SO(5)]$.
As was argued in \ci{mik3} (using a time-averaging procedure 
on the string side),
the quadratic fermionic terms in the two actions 
should indeed match.

Below  we would like to further clarify this
 relation by explaining  the
 truncation of the superstring action that should 
 correspond to a particular fermionic sector on the spin chain side
 and attempting 
   to go beyond  the quadratic level in fermions.

\section{Superstring theory  action}\label{sec3}

One would like  to start with type IIB superstring action  
\ci{MT} and to show that a  fermionic action equivalent
  to the one  found from
the  spin chain in the previous section 
 emerges from it in the ``fast-string'' limit, 
 thus generalizing  the   observations made 
in the bosonic sectors \ci{kru,krt,st,hl1,kt,ital}.
Since the spin chain side is sensitive only to physical degrees of freedom, we are free 
to choose any diffeomorphism and $\kappa$-symmetry gauge. 
We shall use   the string action in the 
light-cone $\kappa$-symmetry gauge of \ci{mt,mtt}, i.e.
  $\Gamma^+ \theta=0$, 
where ``+'' direction is  a light-cone direction in the 
Poincare coordinates of $AdS_5$
space. The advantage of this gauge is that 
the fermionic part of the action (and, in particular, its
 $SU(4)$ structure) becomes relatively simple  and explicit. 

We shall then fix the bosonic conformal gauge and make an ansatz for the 
bosonic $AdS_5$ fields that corresponds to the choice of the 
{\it global}
 $AdS_5$ time being proportional to the 
world-sheet time $\tau$.   
The $t\sim \tau$ relation is needed to  ensure that 
the resulting 2-d Hamiltonian will 
match   the spin chain Hamiltonian 
whose eigen-values should be anomalous dimensions.
 Put differently, 
while we will be  using the $\kappa$-symmetry gauge ``adapted'' to 
the Poincare coordinates, we may still replace the bosonic $AdS_5$ 
Poincare  coordinates by the global $AdS_5$ ones\foot{The standard
transformation is $e^\phi  x^0 = \cosh \rho \ \sin t , \ 
e^\phi  x_i = \sinh  \rho \ n_i  ,\  e^\phi =
\cosh \rho\ \cos t - n_4 \sinh \rho, \  n^2_i + n^2_4=1$.}
and then fix the latter 
(time, radial, and unit 4-vector representing angles of $S^3$)  as 
$t= \nu \tau + ..., \ \rho=0+..., \ 
n_i = 0 + ...$, where dots stand for possible fermionic terms.\foot{Since we 
will be  interested in the $SU(2|3)$ 
sector of states the corresponding string
 states should be rotating in $S^5$
(as in the $SU(3)$ sector),   and,  in addition,  the 
fermionic degrees of freedom
should carry a spin component
in $AdS_5$.} 
The parameter $\nu$ will be related to the (large) angular momentum in $S^5$
and thus $1/\nu$ will be our expansion parameter. 

Next, one  is to try choose  a
 consistent ansatz for the bosonic and fermionic fields 
which would 
restricts the string 
action  to the same sector of states ($SU(2|3)$ or its subsectors)
 that we discussed  above on the spin chain side. 
We shall only consider classical string configurations, 
i.e. semiclassical string states corresponding to coherent states
 of the spin chain; as in the previously discussed bosonic sectors,
 the string  
$\a'$ corrections should correspond to subleading $1/L$ corrections
 on the spin chain
side \ci{ft3,ptt,btz}. 
While the truncation of the  purely-bosonic string 
 sigma model equations 
to particular 
 subsectors is relatively  straightforward \ci{st,kt,ptt},
 this is no longer so in the presence of superstring 
  fermions,
 which,  in particular, couple together the 
$AdS_5$ and $S^5$ parts of the bosonic action. 

One may  try to set part of fermionic fields to zero and make
 certain field redefinitions 
in  order to show the existence of a 
 truncation of    classical string equations 
to a subsector with non-zero fermions.\foot{Consistent truncations
of the {\it phase-space}  equations of the light-cone
superstring in \adss  were recently found in \ci{aat}.
They involve setting to zero certain components 
of the generalized even momenta 
(depending on both bosonic and fermionic
variables)
and are equivalent to the truncations in the Lagrangian 
approach we  discuss
below.}
Having found a consistent truncation 
of the classical string equations, one may then attempt  to 
reconstruct 
(at least in a certain fast-string limit corresponding to
the  leading order 
approximation on the gauge theory side)
 an action that describes  this subsector.

\subsection{
 Superstring Lagrangian in  a light-cone gauge  }

Our starting point will be the
$\k$-symmetry gauge fixed 
 Lagrangian of~\cite{mt}\foot{
We ignore the 
 overall factor of string tension ${\sqrt \lambda \over 2 \pi}=  { R^2
 \ov 2 \pi \a'}$.}
\ba
{\cal L}&=&-\frac{1}{2} \sqrt{g}g^{\mu\nu}
\left[2e^{2\phi}(\p_\mu x^+ \p_\nu x^-+\p_\mu x\p_\nu {\bar x})+ 
\p_\mu\phi\p_\nu\phi+\p_\mu X^M\p_\nu X^M\right]
\nonumber \\ &&
-\frac{i}{2}\sqrt{g}g^{\mu\nu}e^{2\phi}\p_\mu x^+\left[\theta^A\p_\nu\theta_A+\theta_A\p_\nu\theta^A
+\eta^A\p_\nu\eta_A+\eta_A\p_\nu\eta^A\right]\nonumber\\ &&
- i\sqrt{g}g^{\mu\nu}e^{2\phi}\p_\mu x^+ X^N\p_\nu X^M\eta_A\rho^{MN}{}^A{}_B\eta^B\nonumber \\&&
+\ha\sqrt{g}g^{\mu\nu}e^{4\phi}\p_{\mu}x^{+}\p_{\nu}x^{+}\left[
(\eta^{A}\eta_A)^2 +(X^{N}\eta_{A}\rho^{MNA}_{\ \ \ \ \ \ B} \eta^{B})^{2}\right]
\nonumber\\ &&
+\epsilon^{\mu\nu}e^{2\phi}\p_\mu x^+X^M\left(
\eta^A\rho^M_{AB}\p_\nu\theta^B+\eta_A\rho^M{}^{AB}\p_\nu\theta_B\right)
\nonumber \\&&
+i\sqrt{2}\epsilon^{\mu\nu}e^{3\phi}\p_\mu x^+X^M\left(\p_\nu{\bar x}\eta_A\rho^M{}^{AB}\eta_B-
\p_\nu x\eta^A\rho^M{}_{AB}\eta^B\right)\,.\label{su4act}
\ea
Here $\mu,\nu=0,1$ and $\phi, x^0, x^i$ are the Poincare coordinates of $AdS_5$ with 
\be
x^{\pm}=\frac{1}{\sqrt 2}( x^{3}\pm x^{0}) \,,\qquad x=\frac{1}{\sqrt 2}
(x^{1}+ix^{2})\,, \qquad{\bar x}=\frac{1}{\sqrt 2}(x^{1}-ix^{2})\,,
\ee
and $X^M$  ($M,N=1,...,6$) is a unit 6-vector parametrising $S^5$
(the constraint $X^M X^M=1$  can be imposed  with 
 a  Lagrange multiplier $\Lambda$).  
The 4+4 complex Grassmann  fields $\theta_A, \eta_A$ (with
 $A=1,2,3,4$ and  $\theta_A = (\theta^A)^\dagger, \ 
\eta_A = (\eta^A)^\dagger$) 
transform in the fundamental representation of  $SU(4)$.
\foot{We mostly follow the notation of \ci{mt,mtt} but use  $A,B=1,2,3,4$ 
instead of $i,j$ as $SU(4)$ indices.}  
The $4 \times  4$  matrices $\rho^M$ are  
 ``off-diagonal'' blocks of the 
 $SO(6)$ gamma-matrices in the 
chiral  representation (their 
properties  are listed in Appendix A),
and  $\rho^{MN}= -\rho^{[M} \rho^*{}^{N]}$.

Notice that $\theta^A$ enter the action only quadratically
(all quartic fermionic terms involve only $\eta_A$) 
 and thus it could  be
in principle  ``integrated out''. 
 We recall \ci{mt,mtt}
that 
$\theta^A$   correspond to the (linearly realised)  
supersymmetry  generators of the 
superconformal algebra $PSU(4,4|4)$  while $\eta^A$ -- 
to the non-linearly realised 
superconformal generators. 

In what follows we shall choose the conformal gauge for the 2-d 
metric  and  will make the following ansatz for the bosonic $AdS_5$ fields
which corresponds to the global $AdS_5$ time $t= \nu\tau + ..., $ 
namely, 
\be
e^{\phi}=\cos\nu\tau\,,\qquad\! 
x^+=\frac{\tan\nu\tau}{\sqrt{2}}\,,\qquad\!
 x^-=-\frac{\tan\nu\tau}{\sqrt{2}}+f(\tau,\sigma)\,,\qquad\! x={\bar x}=0\,,
\label{adsansatz}
\ee
where $f$ is to be determined. Then 
\be   
e^{2\phi}  \del_0 x^+ = { \nu \over \sqrt 2} \ , \la{ann}  
\ee  
and the $x^+$ equation of motion 
(the one obtained by varying $x^-$) is  automatically satisfied. 
 Since we  would like  also 
to   keep some fermions non-zero,
 it is not {\it a priori}  clear if such an  ansatz is
consistent  with all the equations of motion. 
Indeed, we expect that it will 
place restrictions on  allowed fermions and on $f(\tau,\sigma)$. For
example, setting $x=0$ in  the equation of motion for $x$
is possible  as 
long as $\eta$ satisfies
\be
\p_1(X^M\eta^A\rho^M_{AB}\eta^B)=0\,,\label{xeom}
\ee
plus  a similar  complex conjugate relation
 coming from the ${\bar x}$-equation. 
The $\phi$ equation of motion gives
\ba
\p_0f&=& -iX^N\p_0 X^M\eta_A\rho^{MN}{}^A{}_B\eta^B
-\frac{i}{2}\left(\theta^A\p_0\theta_A+\theta_A\p_0\theta^A
+\eta^A\p_0\eta_A+\eta_A\p_0\eta^A\right)
\nonumber \\&&\qquad
- X^M\left(\eta^A\rho^M_{AB}\p_1\theta^B+\eta_A\rho^{MAB}
\p_1\theta_B\right)\,,\label{fdot}
\ea
while the equation for $x^{-}$  implies 
\ba
\p_1^{2}f&=&\p_{1}\Bigl[-iX^N\p_1 X^M\eta_A\rho^{MN}{}^A{}_B\eta^B
-\frac{i}{2}\left(\theta^A\p_1\theta_A+\theta_A\p_1\theta^A+\eta^A\p_1\eta_A+\eta_A\p_1\eta^A\right)
\nonumber \\&&\qquad
- X^M\left(\eta^A\rho^M_{AB}\p_0\theta^B+\eta_A\rho^{MAB}\p_0\theta_B\right)\Bigr]\,.\label{fpp}
\ea
 The $\theta$ equation of motion  and its conjugate are 
\be
\p_0\theta_A+i\p_1(X^M\rho^M_{AB}\eta^B)=0 \,,\ \ \ \ \ \ 
\p_0\theta^A+i\p_1(X^M\rho^M{}^{AB}\eta_B)=0 
\label{thetaeom} \ea
 These relations may be used to eliminate 
  the $\theta$ fermions
from the action.



The  conformal gauge constraints 
($\frac{\delta S}{\delta g^{\mu\nu}}=0$ with $ \sqrt g g^{\mu \nu} =
\eta^{\mu \nu}$) 
 place further restrictions 
on the allowed fermionic configurations. 
 Using  our ansatz~(\ref{adsansatz}) the 
 one of the two   constraints becomes 
\ba
\nu^2&=&\p_1X^M\p_1X^M+\p_0X^M\p_0X^M+\sqrt{2}\nu{\del_0 f}\nonumber \\&&
+\frac{i\nu}{\sqrt{2}}\left[2X^N\p_0X^M\eta_{A}\rho^{MN}{}^{A}{}_{B}\eta^{B}+
(\theta^{A}\p_0\theta_{A}+\theta_{A}\p_0\theta^{A}
+\eta^{A}\p_0\eta_{A}+\eta_{A}\p_0\eta^{{A}})\right]\nonumber \\
&=&\p_1X^M\p_1X^M+\p_0X^M\p_0X^M
-\sqrt{2}\nu X^M\left(\eta^A\rho^M_{AB}\p_1\theta^B+\eta_A\rho^{MAB}\p_1\theta_B\right)
\label{V1}\,.
\ea
In the last line we have used equation~(\ref{fdot}). 
The other constraint implies 
\ba
0&=&\p_1X^M\p_0X^M+i\frac{\nu}{\sqrt{2}}X^N\p_1X^M\eta_A\rho^{MN}{}^A{}_B\eta^B 
\nonumber \\
  &+& { \nu \ov  \sqrt 2 }  \left[  \del_1 f+ { i \ov 2} 
  ( \theta^A\p_1\theta_A +
\theta_A \p_1\theta^A +\eta^A \p_1\eta_A +\eta_A \p_1\eta^A)
 \right] \,.\label{fp}
\ea
This determines the value of 
 $ \p_1 f$ that should be consistent with 
\rf{fpp}.



\subsection{Fermionic  action in $SU(3)$ notation}
\label{confconst}

As already mentioned, 
we  would like to  consider a subspace of classical 
 string configurations 
that should be dual to spin chain states from $SU(2|3)$ subsector.
The corresponding gauge theory operators are  built out of 3 chiral 
complex combinations of 6 scalars   and the  two 
spinor components of the gluino Weyl fermion.
 The fermions should carry Lorentz spin but should be singlets 
under the Cartan  $[U(1)]^3 $ subgroup of $SO(6)$
whose charges ($S^5$ angular momenta)   are 
carried by the scalars. 
 To identify  the 
corresponding  fermionic components on the string theory side 
we should thus 
do the 3+1 split of the $SU(4)$ fermionic components and at the 
end keep only 
the $SU(3)$ singlet fields.
Thus
a systematic procedure to isolate the $SU(2|3)$ sector
should be based on:  

(i) introducing 3 chiral  bosonic fields $\X_i$ 
and isolating their common  large phase factor $\a$ 
($i=1,2,3$) 
 \bea
&&\rX_i=e^{i\alpha }U_i\,,\ \ \ \ \ \ \ \ \ 
\rX_i\equiv  X_{2i-1}+iX_{2i}, \  \ \ \ \  \ \ \ \   U_i U^*_i =1 \ ,
 \label{rotbos}\\
 &&\a = \nu \tau + v (\tau, \sigma)\ ,  \eea 
 and (ii) 
splitting the $SU(4)$ fermions in \rf{su4act} 
 in 3+1 way 
\be
\eta_A\equiv(\eta_i,\,\eta)\ , \ \ \ \ \ \ \ 
\theta_A \equiv(\theta_i,\,\theta)\,  , \ \ \ \ \ 
\ \ \  i=1,2,3 \ .    \la{spli}
\ee
The   two $SU(3)$ singlet fields $\eta\equiv \eta_4$
and $\theta\equiv \theta_4$ should be eventually related 
to the two fermionic variables $\psi_1,\psi_2$ 
of the spin chain action \rf{six} which are  singlets under $SU(3)$
but are rotated by an additional global  $SU(2)$ symmetry.

Finally, (iii) 
one is  to eliminate $\eta_i$ and $\theta_i$ 
fields from the Lagrangian in the large $\nu$ approximation. 
That step may be facilitated by applying some proper $U(1)$ redefinitions 
of fermions  by $e^{ i  \a}$ factors. 
While such rotations 
 may not be necessary for the ``dummy''  $SU(3)$ 
variables $\eta_i$ and $\theta_i$, we may need them 
in order to relate the singlet fields $\eta,\theta$ 
to $\psi_1,\psi_2$ of the spin chain.

Using the  specific representation of 
 $\rho^M$ matrices and  relations given in Appendix A
  one can  rewrite 
the fermionic part of the  Lagrangian \rf{su4act} 
in the following  manifestly $SU(3)$ invariant form 
depending on  $\X_i, \eta_i, \theta_i, \eta$ and $\theta$
(after using also the ansatz \rf{adsansatz},\rf{ann})
\be \td {\cal L}_F \equiv  \sqrt 2 \nu^{-1} {\cal L}_F
= \td {\cal L}_{2F} + \td {\cal L}_{4F} \ , \ee
where the quadratic terms are
 \be
\td {\cal L}_{2F} &=&i\eta^{i}\p_{0}\eta_{i}+i\etab\p_{0}\eta
+i\theta^{i}\p_{0}\theta_{i}+i\thetab\p_{0}\theta
\nonumber\\
&&+\ \epsilon_{ijk} \eta^{i}\p_{1}\theta^{j}\X^{k}-\epsilon^{ijk}
\eta_{i}\p_{1}\theta_{j}\X_{k}
\nonumber \\ &&
+\ \eta^{i}\p_{1}\thetab\X_{i} -\eta_{i}\p_{1}\theta\X^{i}
+\p_{1}\theta^{i}\etab\X_{i} -\p_{1}\theta_{i}\eta\X^{i} \nonumber \\
&&-\ i (\X^{i}\p_{0} \X_{j}- \X_{j} \p_{0} \X^{i} )\eta_{i} \eta^{j}
-\ i 
\X^{i}\p_{0} \X_{i} (\eta^{j} \eta_{j} - \etab \eta) 
\nonumber \\
&& -\ \ i ( \epsilon^{ijk}\X_{j}\p_{0} \X_{k} \eta_{i} 
\etab  - \epsilon_{ijk}\X^{j}\p_{0} \X^{k} \eta\eta^{i} ) 
\,, \label{su3lag}
\ea
and the quartic terms are
 \ba\la{quat}
\td {\cal L}_{4F} 
= -   {\nu\ov\sqrt 2}  \big(  { 3} \eta^i\eta_i   \etab \eta 
&-& 4 \X_i\eta^i\X^j\eta_j\etab \eta 
+ 4 \eta_i\X^i\eta^j\X_j\eta_k\eta^k  \nonumber \\
&+& 2\epsilon_{ijk}\eta^i\eta^j\X^k\eta_l\X^l\eta 
 +  2\epsilon^{ijk}\eta_i\eta_j\X_k\eta^l\X_l\etab \big) \ ,  
\ea
where 
\be \X^i = \X_i^*\ ,\ \ \ \ \ \  \  \eta^i= \eta_i^\dagger\ , \ \ \ \ \ \ 
\theta^i= \theta_i^\dagger\ , \ \ \ \ \ \ \ \   \bar \eta = \eta^\dagger\ , \ 
\ \ \ \ \ \ \bar \theta = \theta^\dagger\ . \ee 
For completeness, the bosonic $S^5$ part of the  Lagrangian 
\rf{su4act}  written in terms of $\X_i$ is 
\be \la{bbo}
{\cal L}_{B} = - \ha \del^\m \X^*_i \del_\m \X_i
+ \ha  \Lambda (  \X^*_i  \X_i -1 ) \ . \ee

\section{Some  fermionic  solutions to superstring
 equations of motion }\label{sec4}

In this section we present a number of 
simple classical rotating string solutions of the above action 
that have  non-zero fermions. 
This will help to understand better which truncations 
of the superstring coordinates are consistent with equations of motion. 
The solutions we shall discuss  are 
generalisations of the rotating 
 circular string 
solutions found in \ci{ft2,art}. 

Our starting point  will be the 
 action~(\ref{su3lag}),\rf{quat}. 
 There are a number of consistent truncations of the
  Lagrangian~(\ref{su3lag}). 
  One  includes restricting the bosonic fields 
  to $AdS_3$ inside $AdS_5$ and $S^3$  inside $S^5$
  and also truncating the fermions in one of two possible ways, i.e. 
\be
 (x,\X_3; \eta ,\eta_3, \theta_1 ,  \theta_2 
 )=0\,,\ \ \ \  {\rm i.e.} \ \ \ \ 
   (\X_1,\X_2; \theta, \theta_3,\eta_1, \eta_2 ) \not=0\ , 
 \label{su12theta}\ee
 or \be 
 ( x,\X_3;  \theta , \theta_3,  \eta_1 , \eta_2 ) =0 \,,
 \ \ \ \ 
 {\rm i.e.} \ \ \ \ 
   (\X_1,\X_2;\eta,\eta_3,\theta_1, \theta_2) \not=0\ . \label{su12eta}
\ee
We can also 
 restrict to $AdS_3$ inside $AdS_5$ and $S^1$ inside $S^5$
 and truncate fermions further in one of the two ways 
\be
 (x,\X_2,\X_3; \eta ,\eta_2,\eta_3, \theta_1 ,  \theta_2, \theta_3)=0\,,\ \ \ \  {\rm i.e.} \ \ \ \ 
   (\X_1; \theta, \eta_1 ) \not=0\ , 
 \label{psu11theta}\ee
 or 
 \be 
 ( x,\X_2,\X_3;  \theta ,\theta_2 ,   \theta_3 , 
    \eta_1, \eta_2 , \eta_3) =0 \,,
 \ \ \ \ 
 {\rm i.e.} \ \ \ \ 
   (\X_1;\eta, \theta_1) \not=0\ . \label{psu11eta}
\ee
It is natural to expect that 
after  integrating out ``extra''   
fermions (i.e. leaving only $\eta$ or $\theta$ 
in each case) 
these  subsectors may be related to 
the $SU(1|2)$  and $SU(1|1)$
gauge theory sectors. In this section 
we shall use the names  ``$SU(1|2)$''  and ``$SU(1|1)$''
   for the above  superstring truncations. 
Similar truncations were  also
 obtained in~\cite{aat} using  phase space 
 formulation.



\subsection{``$SU(1|1)$'' fermionic string 
 solution}\label{sec41}

Below we present a particular ``circular''  
string solution for the ansatz~(\ref{psu11eta}).
 We shall take the $AdS_5$ fields to be of the form
  given in equation~(\ref{adsansatz}).
   The
$\eta$ equation of motion then reduces to
\be
0=\p_0^2\X_1\eta-\X_1\p_0^2\eta+\p_1^2(\X_1\eta)\,.\label{etaeom}
\ee
We will solve this by taking  ($|\X_{1}|^{2}=1$)
\be
\X_1=e^{i\nu\tau}(1-iC\tau\bzz)\,,
\qquad\eta=e^{i(n\sigma+\omega \tau)}\z \,,\qquad
\theta=e^{i(n\sigma+(\omega+\nu) \tau)}\bz\,,
\ee
where $\z$ is a constant complex Grassmann number, 
 $C$ is a real constant and
 \be
 \omega=\sqrt{n^2+\nu^2}\,.
 \ee
The $\theta_1$ equation of motion~(\ref{thetaeom}) then gives
\be
\theta_1 =-i\frac{\omega+\nu}{n}e^{-i(n\sigma+
(\omega-\nu)\tau)}\bz 
=-i\frac{\omega+\nu}{n}e^{i\nu\tau}\etab\,,
\ee
while the $\eta_1$ equation implies that
\be
\eta_1 =i\frac{\omega+\nu}{n}e^{-i(n\sigma+
\omega\tau)}\z 
=i\frac{\omega+\nu}{n}e^{i\nu\tau}\thetab\,.
\ee

\noindent The $\X_{1}$ equation of motion is satisfied then if the 
 Lagrange multiplier is  
\be
\Lambda=-\nu^2- A\bzz\,,\ \ \ \ \ \ \ \ 
A=-2\sqrt{2}\nu^{2}-\frac{2\sqrt{2}\nu^3}{n^2}(\nu-\omega_{n})-2\nu C\,.
\ee
The  $\phi$ equation of motion gives
$\p_0 f=0$, while the conformal gauge constraint~(\ref{fp})
 implies that $\p_1 f=0$. 
In other words,  this solution has the {\it same} $AdS_5$ 
part  as the bosonic solutions representing 
 strings rotating on $S^5$.
It is  easy to see that eq.(\ref{fpp}) is also satisfied.
Finally, the conformal gauge constraint~(\ref{V1}) is satisfied for
\be
C=-2\sqrt{2}(\omega-\nu)\,.
\ee
The solution has energy $\nu$, and is charged under the Cartan
generators $J^A{}_A$ of $SU(4)$, with all other $SU(4)$ charges zero. 
Indeed, for this solution $J^1{}_1=-J^2{}_2=-J^3{}_3=J^4{}_4\equiv J$, where
\be
J=-\frac{1}{2}\nu+\sqrt{2}\bzz\left(\omega-\nu-\frac{\omega^2-\nu\omega}{2m^2}\right)\,.\label{jpsu11}
\ee
We have thus obtained a formal 
classical superstring solution 
which generalizes 
the BMN  geodesic  solution ($t= \nu \tau, \ \X_1 = e^{i \nu \tau}$)
to the presence of  non-trivial $\s$-dependent 
 fermions. Here the string is ``spread'' only in the odd 
 directions of superspace and in the even $x^-$ direction. Its charge 
depends on $\bzz$, and hence appears to be Grassmann valued.
This is an artifact of our semi-classical treatment of fermions;
one  may view  $\bzz$ in equation~(\ref{jpsu11}) as a 
real-valued expectation value
$\left<\bzz\right>$.

\subsection{``$SU(1|2)$'' solution with $\theta\not=0$} \label{sec42}

Let us now  present a solution in the case of  the
 truncation~(\ref{su12theta}). Guided by analogy with the solution 
of the  \LL  model 
 in 
section~\ref{gaugesoln}, we will try the following ansatz
\ba
\X_1&=&\frac{1}{\sqrt{2}}\left(e^{in\sigma}-  e^{-in\sigma}\bzz\right)
e^{i\ww\tau -F(\tau,\sigma)\bzz- iG(\tau,\sigma)\bzz}\,,\label{X1ansatz}\\
\X_2&=&\frac{1}{\sqrt{2}}\left(e^{-in\sigma}+
 e^{in\sigma}\bzz \right)e^{i\ww\tau +F(\tau,\sigma)\bzz- iG(\tau,\sigma)\bzz}\,,\label{X2ansatz}\\
\theta&=&e^{im\sigma+i\omega\tau}\z\label{etaansatz}\,,
\ea
together with $\theta_3=0$ and
\ba
\eta_1&=&\frac{A_1}{\sqrt{2}}e^{i((n-m)\sigma+(\ww-\omega)\tau)}\bz\,,\label{eta1ansatz}\\
\eta_2&=&\frac{A_2}{\sqrt{2}}e^{i(-(n+m)\sigma+(\ww-\omega)\tau)}\bz\,.\label{eta2ansatz}
\ea
As above,  $\z$ is a constant complex Grassmann parameter, 
$G$ and $F$ are real  $\sigma$-periodic function and the 
$A_i$ are constants.\foot{Of course, 
$  e^{ -iG\bzz}= 1 - iG \bzz$
but we prefer to use the exponential parametrization.}
   The equations of
    motion
    for 
    $\eta$, $\theta_1$, $\theta_2$ and $\eta_3$  are then trivially 
satisfied. The $\theta$, $\eta_1$, $\eta_2$ and $\theta_3$ equations of
 motion reduce to the following constraints
\ba
0&=&A_1-A_2\,,\label{PP1}\\
0&=&(A_1+A_2)m-2i\omega\,,\label{PP2}\\
0&=&A_1(\omega-\ww)-A_2\ww-im\,,\label{PP3}\\
0&=&A_2(\omega-\ww)-A_1\ww-im\label{PP4}\,.
\ea
The solution to these is 
\be
\omega=\ww\pm \sqrt{\ww^2+m^2}\,,\qquad A_1=A_2=-i\frac{\ww\pm\sqrt{\ww^2+m^2}}{m}\,.
\ee
Turning to the bosonic equations of motion it is easy to see that 
the $\X_3$ equation of motion is trivial while the $\X_1,\X_2$ ones 
 reduce to the condition
\be
0=4n\p_1G-2\p_1^2F+4i\ww\p_0F+2\p_0^2F
\,.\label{CC1}
\ee
The  equations of motion for the $AdS_5$ coordinates 
and the conformal gauge constraints give rise to 
further constraints 
\ba
0&=&\p_0\p_1G+2n\p_0F\,,\label{CC2}\\
0&=&\sqrt{2}(\ww\pm\sqrt{m^2+\ww^2})\nu-\ww\p_0G\,,\label{CC3}\\
0&=&\p_1^2G+2n\p_1F\,,\label{CC4}
\ea
as well as
\be
\ww=\sqrt{\nu^2-n^2}\,.
\ee
Given the equations~(\ref{PP1})--(\ref{PP4}), the 
equation of motion for $\phi$~(\ref{fdot}) implies that
\be
\p_0f=0\,,
\ee
while the conformal gauge constraint~(\ref{fp}) reduces to
\be
\p_1f=\frac{1}{m\nu}\left(2(m^2+\ww^2\pm \ww\sqrt{m^2+\ww^2})
\nu-2\sqrt{2}mn\ww F-\sqrt{2}m\ww\p_1G\right)\bzz\,.
\ee
The simplest solution to these equations which ensures that $x^+=-x^-$ is
\ba
F(\tau,\sigma)&=&-\frac{\sqrt{2}(m^2+\ww^2\pm \ww\sqrt{m^2+
\ww^2})\nu}{2mn\ww}\,,\\
G(\tau,\sigma)&=&\sqrt{2}\nu\  [\ 1\pm\sqrt{1+(m/\ww)^2}\ ]\ \tau\,.
\ee
 One can  check that for this solution the 
 Cartan charges $J^A{}_A$ are the only non-zero components of the $SU(4)$ charges (see Appendix B).

 To match this solution
  to the spin-chain sigma model one, we need to 
  take $\nu\rightarrow 0$. 
In order to keep our solution finite in this limit
 we will consider the minus sign choice 
 in the above relations. In this limit $A_1$ and $A_2$ tend to 
 zero, and, 
as a result,  the only non-zero fields are $\X_1$, $\X_2$ and $\theta$, which can be matched to the spin chain variables. 
We should  stress, however, 
 that besides being rotated by a common 
 phase $G\bzz$, the $\X_i$ are also rescaled by
 factor  $(1\pm F\bzz)$. 
This implies that while a rotation  by a common phase, 
discussed below in section~5, can be used to 
relate the string and spin chain variables and 
actions to the leading order,
 at higher orders one will need more involved field
redefinitions.


\subsection{``$SU(1|2)$'' solution with 
 $\eta\not=0$}

Let us now  consider the case of the truncation~(\ref{su12eta}). In this sector it turns out 
that one needs to consider a more general
  ansatz for the bosons
\ba
\X_1&=&\frac{1}{\sqrt{2}}\left(e^{in\sigma}-
 e^{-in\sigma}\bzz\right)e^{i\ww\tau-F(\tau,\sigma)\bzz - iG(\tau,\sigma)\bzz}\,,
\label{X11ansatz}\\
\X_2&=&\frac{1}{\sqrt{2}}\left(e^{-in\sigma}+ 
e^{in\sigma}\bzz\right)e^{i\ww\tau+F(\tau,\sigma)\bzz- iH
(\tau,\sigma)\bzz}\,,\label{X12ansatz}
\ea  
where  $F$,  $G$ and $H$  are 
  real $\sigma$-periodic functions. 
 For the fermions we shall choose
 \bea
\eta&=&e^{im\sigma+i\omega\tau}\z\,,\qquad\qquad\qquad\,\,\,\,\,\,
\eta_3=Be^{i(m\sigma+(\omega-2\ww)\tau)}\z\,,
\label{etaiansatz}\\
\theta_1&=&\frac{A_1}{\sqrt{2}}e^{i((n-m)\sigma+(\ww-\omega)\tau)}\bz\,,\qquad
\theta_2=\frac{A_2}{\sqrt{2}}e^{i((-n-m)\sigma+
(\ww-\omega)\tau)}\bz\,.\label{thansatz}
\eea
 The fermionic equations of motion then reduce to 
\ba
0&=&A_1(m-n)+A_2(m+n)+2i(\ww+\om)\,,\label{P1}\\
0&=&(m-n)(1-B^*)+iA_1(\ww-\om)\,,\label{P2}\\
0&=&(m+n)(1+B^*)+iA_2(\ww-\om)\,,\label{P3}\\
0&=&A_1(m-n)-A_2(m+n)+2iB^*(3\ww-\om)\label{P4}\,.
\ea
These equations can be solved for $A_i,\,B$ and $\om$ in terms of
 $n,\,m$ and $\ww$. 
The general solution is quite 
involved, but setting $n=m$ gives  three simple solutions 
\ba
I: \ \ \  \om&=&\ww\,,\qquad A_1=\mbox{free}\,,\qquad\,\,
A_2=-\frac{2i\ww}{n}\,,\qquad B=-1\,,\\
II_{\pm}:\ \ \ \om&=&\ww\pm 2\sqrt{\ww^2+ n^2 }\,,\qquad\qquad\,\, 
A_1=0\,,\qquad \nonumber \\
A_2&=&\frac{2(-i\ww\mp i\sqrt{\ww^2+n^2})}{n}\,,\ \ \ 
B=\frac{n^2+2(\ww^2\pm \ww\sqrt{\ww^2+ n^2})}{n^2}.
\ea
\commentout{
To simplify matters, let us 
 require that the off-diagonal components of the  $SU(4)$ 
 charge matrix  vanish for our solution. 
The  components that do not automatically vanish are 
\ba
J^3{}_4&=&-\frac{1}{4}e^{2i\ww\tau}\bzz  \int
d\sigma\left[\p_0N-4\sqrt{2}\nu B^*\right]\,, \\
J^4{}_3&=&\frac{1}{4}e^{-2i\ww\tau}\bzz  \int
d\sigma\left[\p_0N-4\sqrt{2}\nu B\right]\,,\\
J^1{}_2&=-&\frac{i}{4}\bzz   \int
d\sigma e^{-2in\sigma}\left[2\ww N+i\p_0M\right]\,,\\
J^2{}_1&=& \frac{i}{4}\bzz  \int
d\sigma e^{2in\sigma}\left[2\ww N-i\p_0M\right]\,,
\ea
where we have defined
\be
M(\tau,\sigma)=G(\tau,\sigma)+H(\tau,\sigma)\,,\qquad
N(\tau,\sigma)=G(\tau,\sigma)-H(\tau,\sigma)\,.
\ee
The simplest solution to  $J^3{}_4=0, \ J^4{}_3=0$  is
\be
N=4\sqrt{2}\nu B^*\tau + N_0\,,\ \ \ \ \  \ \ \ \ \  N_0 =\const \ . 
\ee
}
For the solution $II_-$, the $\X_i$ equations of motion reduce to 
\ba
0&=&8\sqrt{2}\ww\nu+\frac{16\sqrt{2}\nu \ww^2(\ww-\nu)}{n^2}
+2n\p_1M-2\p_1^2F-i\p_1^2N\nonumber \\&&
+4i\ww\p_0F-2w\p_0N+2\p_0^2F+i\p_0^2N\,,
\ea
where we have defined
\be
M(\tau,\sigma)=G(\tau,\sigma)+H(\tau,\sigma)\,,\qquad
N(\tau,\sigma)=G(\tau,\sigma)-H(\tau,\sigma)\,.
\ee
The $AdS_5$ equations of motion and the conformal gauge 
constraints reduce to
\ba
0&=&4n\ww\p_1F+\ww\p_1^2M+n\p_0^2M\,,\\
0&=&4n\ww\p_0F+\ww\p_0\p_1M+n\p_1^2M\,,\\
0&=&8\sqrt{2}\ww\nu(n^2+2\ww^2-2\ww\nu)+n^3\p_1N+n^2\ww\p_0M\,,
\ea
together with the condition
\be
w=\sqrt{\nu^2-n^2}\,,
\ee
and the following equations for $f$
\ba
\p_0f&=&0\,,\\
\p_1f&=&-8\nu^3(n^2-2\nu^2+2\nu \ww)+\sqrt{2}n^3\left(4n\ww F+\ww\p_1M+n\p_0N\right)\,.
\ea
A simple solution to these equations is 
\ba
M&=&-\frac{8\sqrt{2}\nu(n^2+2\ww^2-2\ww\nu)}{n^2}\tau\,,\\
N&=&\frac{4\sqrt{2}\nu( n^2+2 \ww^2-2\ww\nu)}{n^2}\tau\,,\\
F&=&\frac{\sqrt{2}\nu(n^2+\nu^2)( n^2+2 \ww^2-2\ww\nu)}{wn^4}\,,
\ea
 and $f=0$ which implies 
  $x^+=-x^-$. While the existence 
   of this exact solution is quite remarkable, 
we should stress that its complexity 
(in particular, the fact that the 
phases of $\X_1$ and $\X_2$ are different)
indicates again 
the need for some further field redefinitions to
match the string and  the spin chain variables. 
This solution also shows 
the difference between the $\eta=\eta_4$ and $\theta=\theta_4$
 subsectors on the string side.
 Comparing  to  
the solution  in the previous subsection it is 
 clear that some field redefinitions are needed 
 to make explicit the $SU(2)$ symmetry
 between the two fermions $\theta$ and $\eta$.

\commentout{
Using eqs.(\ref{P1}) 
and~(\ref{P4}) this gives 
\be
0=\p_1M\,, \ \ \ \ \ {\rm i.e.}   \ \ \ \ \ \ \ M=  M(\tau)
\ . \label{C1}
\ea
 The remaining bosonic equations of motion, 
together with the conformal gauge constraints  require
further  that
\ba
0&=&\ww\p_0\p_1M+n\p_0^2N\,.\label{C2}\\
0&=&\ww\p^2_1M+n\p_0\p_1N\,.\label{C4}\\
0&=&-4\sqrt{2}\nu\left[\ww+\omega+|B|^2(3\ww-\omega)\right]
+2n\p_1N+2\ww\p_0M\,, \label{C3}
\ea
as well as
\be
\ww^2=\nu^2-n^2\,.
\ee
These are easily solved with 
\be
M=2\frac{\sqrt{2}\nu}{\ww}\left[\ww+\omega+|B|^2(3\ww-\omega)\right]\tau+
M_0\,,\ \ \ \ \    M_0=\const \ . 
\ee}


\section{Matching the string and spin-chain actions}\label{sec5}

Let us now discuss how to relate  the  \LL action \rf{su31} 
representing the low-energy  coherent states of the $SU(2|3)$
spin chain  to a ``fast-string'' limit 
of the superstring action 
\rf{su4act}  or \rf{su3lag},\rf{quat},\rf{bbo}.
We shall first consider the quadratic fermionic term in the general 
$SU(2|3)$ case (to ``one-loop'' or leading term 
in ``fast-string'' expansion)   and then discuss  the special 
case of $SU(1|1)$ sector (including also subleading terms).

\subsection{$SU(2|3)$ case to leading order }

As was mentioned 
 in sect. 3.2, we should isolate the common large phase 
$\a$ of
the $S^5$  bosons 
as in \rf{rotbos}
and simplify the Lagrangian  assuming that $\nu$ in  $\a = \nu \tau + v
$ large.
In order to do that one   may  also    redefine the 
two pairs of 3+1 fermions 
as follows
\ba
\eta_{i}\to \frac{1}{\nu}e^{i\nu\tau}\xi_{i}\ , \ \ \ \ \ \ 
\eta\to e^{-i\nu\tau}\psi\ ,  \ \ \ \ \ 
\theta_{i} \to \frac{1}{\nu}e^{2i\nu\tau}\zeta_{i}\ ,  \ \ \ \ \ 
\theta \to  \theta\,, \label{fermrot}
\ea
where in general $\nu \tau$ should be replaced by $\a= \nu \tau + v$.
Note that after  these rotations the 
$\epsilon_{ijk} \eta^i \p_1 \theta^j \X^k$ terms  in \rf{su3lag} 
have large  non-vanishing phases and thus average to zero as in 
\ci{mik12,mik3,kt}.\foot{A possible alternative to using the averaging
procedure may be to  splitting $\eta_i$ into the 
transverse and longitudinal part with respect to $U^i$ 
$ 
\eta_i = \etap_i  + U_i q , \    q= U^i  \eta_i , \ \ 
 U^i \etap_i =0$
 and to try to decouple $q$ and  $\etap_i$.}
 The  remaining terms in 
\rf{su3lag} become  ($U^i \equiv U_i^*$)
\ba
\td {\cal L}_{2F}&=&i\psib\p_{0}\psi
+i\thetab\p_{0}\theta 
+U_{i}\xi^{i}\p_{1}\thetab-U^{i}\xi_{i}\p_{1}\theta
+U_{i}\p_{1}\zeta^{i}\psib-U^{i}\p_{1}\zeta_{i}\psi\nonumber \\
&& +\ 2\zeta_{i}\zeta^{i} +   2U^{i}U_{j}\xi_{i} \xi^{j} + 
 i U^{i}\p_{0}U_{i}\psib \psi 
\label{rotsu3lag} \\
&&
+\ \frac{i}{\nu^{2}}\Bigl[\xi^{i}\p_{0}\xi_{i}+
\zeta^{i}\p_{0}\zeta_{i}
-(U^{i}\p_{0} U_{j}- U_{j} \p_{0} U^{i})\xi_{i} \xi^{j}
-  U^{i}\p_{0} U_{i}\xi^{j}\xi_{j}\Bigr]\,.\nonumber 
\ea
Dropping  the subleading $1\ov \nu^2$ term  we observe that $U^i \xi_i$
and  $\zeta_i$  become non-dynamical variables 
 that can be easily
 solved for and then eliminated from the Lagrangian:
\ba\la{use}
 U^i \xi_{i}= \frac{1}{2} \p_{1}\thetab\ ,
 \ \ \ \ \ \ \ \ \ \ \
\zeta_{i}= -\frac{1}{2}\p_{1}(U_{i}\psib)\,. 
\ea
Then, to leading order in $\nu$, 
the quadratic part of the Lagrangian~(\ref{rotsu3lag}) is 
\be
\td {\cal L}_{2F}=i\psib\p_{0}\psi
+i\thetab\p_{0}\theta-  \frac{1}{2}\p_{1}\thetab \p_{1}\theta 
- \frac{1}{2}\p_1(U_{i}\psib) \p_1(U^{i}\psi) 
+ i U^{i}\p_{0}U_{i} \psib \psi \,.
\label{su32befrot}
\ee
Solving  the 
conformal gauge constraints 
(\ref{V1}),\rf{fp} 
we  obtain~\footnote{We
 use the identity $U^iD_1U_i=0$. We  also drop by 
``averaging'' the same terms that we omitted  in getting
 from eq.~(\ref{su3lag}) to eq.~(\ref{rotsu3lag}). 
 Equivalently, these 
terms should
  be dropped 
already in the action (\ref{su4act}).}
\ba
\p_{0}v&=&-C_{0}-\frac{1}{2}|D_{1}U_{i}|^{2}+\dots
\,, \qquad 
\p_{1}v=-C_{1} + \dots\,,\qquad \!\!C_a = -i U^i \del_a U_i \,. 
\label{pdofu} \ea
These relations will be modified by fermionic terms
 indicated by $\dots$. To determine the quadratic 
term in the Lagrangian it is, however, enough 
to ignore these terms.
In order to match the spin-chain action \rf{su31} 
for $(U_i,\psi_\alpha)$
we need an extra  redefinition of   the fermions
$\theta$ and $\psi$ 
\be
(\theta, \psi) \rightarrow 2^{1/4}e^{iv} (\psi_1\,,\psi_2\,) \ .
\ee
Then the  Lagrangian~(\ref{su32befrot}) takes the form 
\be
\td {\cal L}_{2F}=i\bar \psi_\a D_{0}\psi_\a  + 
\frac{1}{2} |D_{1}U_{i}|^2  \bar \psi_\a \psi_\a
-  \frac{1}{2}D_{1}^{*}\bar \psi_\a D_{1}\psi_\a   \ . 
\label{gsu31}
\ee
Combined with the $SU(3)$ sector  bosonic contribution 
\ci{st} from \rf{bb} this 
 is almost identical to the quadratic  part of the spin 
chain Lagrangian~(\ref{su31}) apart from the  minus sign
in the fermionic $D_0$  term. This sign can be matched  
  by renaming $ \tau \to - \tau$ and 
$U_i\to U^*_i$  in relating  the string action  to 
the spin chain action.

\bigskip

What remains is to show  that
(i)  the  ansatz \rf{adsansatz}
for the bosonic  we used is consistent, 
and (ii)  quartic fermionic terms also match. 
To demonstrate (i) 
 one is to show, in particular,  that  the two equations for $x^-$ 
implied by  the $\phi$ and $x^{+}$ equations of motion
following from \rf{su4act} or \rf{su3lag} are indeed
consistent with each another, and that the two equations for $v$ 
given in~(\ref{pdofu}) are also consistent. Since 
we have only worked to quadratic order these questions can be 
justifiably ignored in our treatment, but need to be addressed as part of understanding 
(ii).

Proving (ii)  may involve an additional  
field redefinition which we did not find. 
We shall only mention that the structure of
 the quartic terms in \rf{quat} 
(where the $\epsilon_{ijk}$ terms should not 
contribute after time averaging)
is, in principle, consistent with that 
of quartic fermionic terms in \rf{su31}
(which contain spatial  derivatives of fermions) 
after one uses \rf{fermrot},\rf{use} 
 to eliminate $U^i \xi_i$ in terms of $\del_1 \theta$.

Let us now  look at   subsectors 
of the superstring action  and discuss 
their relation to the corresponding  subsectors of the \LL 
action.

\subsection{ $SU(1|1)$ case }
\label{cont}

Let us  specialize to the 
 $SU(1|1)$ sector where  $U_1=1, \ U_2,U_3=0$
and there is just one  fermionic degree of freedom.
The quadratic fermionic part of 
both the spin chain \rf{su31}  and the 
string \rf{gsu31}  Lagrangians  reduces to the 
following  leading-order 
term (here we rescale $\tau \to t$ and 
set $\td \l = {\l \ov J^2}$) 
\be 
 {\cal L}=  - i \psib \del_t \psi  - 
\ha  \td \l\  \del_1 \psib  \del_1  \psi  \ .  \la{lead} \ee
This  may be interpreted as a Lagrangian 
for  a non-relativistic fermion
(see also \ci{hl1}).\foot{The fact that a massive 
non-relativistic fermion 
action appears in the coherent state 
path integral of the   XY spin chain in a magnetic field 
is well-known \ci{sac}. For a fine-tuned coefficient 
of the magnetic field the XY model has \ci{staud} 
hidden $SU(1|1)$ symmetry (a relation of this spin chain to free 
fermion was pointed out earlier in \ci{calw}).}

One interesting  question is how that action extends to higher orders 
in $ \td \l$. The  answer turns out to be  that
 it is  just 
a natural   ``relativistic'' generalization (up to
 integration by parts):
\be 
{\cal L} =  - i \psib \del_t \psi  -  
 \psib  \left(\sqrt{ 1 - \tl \del^2_1 } - 1 \right)   \psi  \ . \la{rela}  \ee
This  the action that reproduces the  equation 
for the upper  component of the 2-d massive 2-component Dirac fermion
(upon elimination of the other component) 
with mass $m = { 1\ov  \sqrt
\tl}= {J \ov \sqrt \l}$;
 it is thus in agreement with the 
BMN spectrum to all orders in $\tl$. 
This action is in agreement with 
 recent results  on the higher-order generalization of the
Bethe ansatz in the $SU(1|1)$ sector:
the above expression 
(or its direct discretization)
 reproduces the leading ``BMN'' terms in the corresponding Bethe 
ansatz expressions in \ci{calw,staud}.

The bosonic analog of \rf{rela} 
 appeared already in the discussion of the  $SU(2)$ sector in 
\ci{krt}: there the sum 
 of  all terms in the 
 string effective
 Hamiltonian  that are of second order  in the 3-vector $\vn$
 parametrising the semiclassical  state of a fast string 
 (or coherent state of the spin chain ) 
 was found to be 
 \be \la{gy}
{\cal L} =  \vec C(n)  \del_t   \dot{  \vec n } - { 1 \ov 4} \vn \left( \sqrt{1 - \tl\ 
\del^2} -1\right)  \vn  +  O(\vn^4) \ . \ee
This expression is  in agreement with the few leading-order
 results for the coherent-state action derived from the 
 $SU(2)$  sector dilatation operator  
and with the exact BMN  spectrum \ci{ryz}.\foot{The  
coherent state analogs of the BMN states are  
 small fluctuations near the vacuum state $\vn_0 = (0,0,1)$. 
 On the spin chain side these correspond (in the discrete version) 
 to the microscopic spin wave excitations or magnons. Similar 
 relation appears in the $SU(1|1)$ sector \ci{calw,staud}.}

Let us explain now 
how  the  action \rf{rela} can  be derived from the full  superstring 
Lagrangian (we shall consider only the quadratic terms in the latter). 
One  expects  to reproduce the  BMN-type massive fermion
action for quadratic fermionic fluctuations  in the case 
of the point-like bosonic  background 
\be \la{red} \X_1 = e^{i \a}\ , \ \ \ \ \ \ \ \ \ \X_2,\X_3=0 \ . \ee
Using \rf{red} in the action \rf{su3lag}
we get (here $a,b=2,3$) 
\ba
\td {\cal L}_F&=&i\eta^{1}\p_{0}\eta_{1}+ i\eta^{a}\p_{0}\eta_{a}
+   i\etab\p_{0}\eta
+i\theta^{1}\p_{0}\theta_{1 }+  i\theta^{a}\p_{0}\theta_{a } 
+  i\thetab\p_{0}\theta\nonumber\\
&&+\ \epsilon_{ab} \eta^{a}\p_{1}\theta^{b}e^{- i \a}    -\epsilon^{ab}
\eta_{a}\p_{1}\theta_{b}  e^{i \a} 
+\eta^{1}\p_{1}\thetab e^{i \a} -\eta_{1}\p_{1}\theta e^{- i \a} 
+\p_{1}\theta^{1}\etab e^{i \a}  -\p_{1}\theta_{1}\eta e^{- i \a}  \nonumber \\
&& -   \del_0 \a  ( \eta^{1} \eta_{1}   +   \etab \eta - \eta^{a} \eta_{a}  )  
+ O(\eta^4) 
\,. \label{su1}
\ea
It is clear now that $\theta_a$ and $\eta_a$ decouple from the singlet sector and
we can consistently set them to zero. The same conclusion remains 
after we include the quartic fermionic term \rf{quat}
which reduces simply to $ {\nu \ov \sqrt2} \eta^1 \eta_1 \bar \eta \eta$. 
Then we are left with 
\ba
\td {\cal L}_F&=&i\eta^{1}\p_{0}\eta_{1}
+   i\etab\p_{0}\eta
+i\theta^{1}\p_{0}\theta_{1 } +  i\thetab\p_{0}\theta\nonumber\\
&&
+\ \eta^{1}\p_{1}\thetab e^{i \a} -\eta_{1}\p_{1}\theta e^{- i \a} 
+\p_{1}\theta^{1}\etab e^{i \a}  -\p_{1}\theta_{1}\eta e^{- i \a}  \nonumber \\
&& -   \del_0 \a  ( \eta^{1} \eta_{1}   +   \etab \eta   )  +  
{\nu \ov \sqrt2} \eta^1 \eta_1 \bar \eta \eta
\,.  \label{sus}
\ea
Now we can further do one of the two possible  truncations
(or  \rf{su12eta},\rf{su12theta})
$$
 \ \ \ \theta_1=\eta=0\ \ \ \  { or} \ \ \ \ \ \ 
  \ \ \ \eta_1=\theta=0  \ . $$
Both are consistent choices, and 
in both cases the quartic fermionic term  vanishes. 
In the first case we end up with 
\ba
\td {\cal L}_F= i\eta^{1}\p_{0}\eta_{1}
 +  i\thetab\p_{0}\theta
+ \eta^{1}\p_{1}\thetab e^{i \a} -\eta_{1}\p_{1}\theta e^{- i \a} 
  -   \del_0 \a   \eta^{1} \eta_{1}  
\,,   \label{usus}
\ea
while in the second 
\ba
\td {\cal L}_F=   i\etab\p_{0}\eta
+i\theta^{1}\p_{0}\theta_{1 } 
+\p_{1}\theta^{1}\etab e^{i \a}  -\p_{1}\theta_{1}\eta e^{- i \a}  -  
 \del_0 \a     \etab \eta  
\,.    \label{bus}
\ea
What remains then  is to  integrate out 
$\eta_1$ in the first case or $\theta_1$ in the second.

More precisely, one should  ensure that the remaining singlet
fields are kept ``massless''  and  eliminate  time-dependent 
exponential factors in the mixing terms. That means that 
 in the first 
case  one should first apply  the same  redefinition
as in \rf{fermrot}, i.e. 
$ \eta_1 \to e^{i \a} \eta_1$, $\theta \to \theta$. 
Then the mass of $\eta_1$ doubles,   and  integrating it out we get
\ba
\td {\cal L}_F= 
   i\thetab\p_{0}\theta  -   
   \p_{1}\thetab { 1 \ov  2 \nu  - i \del_0 }    \p_1 \theta     
 =    i\thetab\p_{0}\theta  -   
  { 1 \ov  2 \nu   } \p_{1}\thetab     \p_1 \theta    
  + ...                     \,.   \label{lus}
\ea
In the second case,  we should 
 keep $\eta$ as  a  ``light'' field and so 
should do a redefinition to absorb its mass term
$ \eta \to e^{-i \a} \eta$, and then do a compensating
 redefinition of 
$\theta_1 \to e^{2 i \a} \theta_1$
 to eliminate the exponential phase  factors in the
mixing terms. The resulting  redefinition is then the same as in 
\rf{fermrot}.
 as a result, 
  we end up with the same 
 action for redefined $(\eta,\theta_1) \equiv 
 (\psi,\zeta)$ as  for the redefined  $(\theta,\eta_1)
 \equiv  (\theta, \xi) $ in the
 first case, i.e. 
 \ba
\td {\cal L} =   i\psib \p_{0}\psi +i\zb\p_{0}\z  - 2 \nu \zb \z  
+\p_{1} \zb\ \psib   - \p_{1}\z\ \psi  
\,.     \label{pus}
\ea
Eliminating the massive $\z$ field, we finish with the same 
action as in \rf{lus}
\ba
\td {\cal L} =   i\psib \p_{0}\psi -  \p_{1}\psib  { 1 \ov  2 \nu  - i \del_0 }    \p_1 \psi
\,.  \label{kus}
\ea
 This provides a   justification for the redefinition used in 
 \rf{fermrot}.
 
 The second term in  \rf{kus} should  be treated perturbatively
 in $\del_0/\nu$ (assuming the  large $\n$ limit). 
 An  equivalent action that leads to the same equations 
 of motion is then 
 \ba
\td {\cal L} =   i\psib \p_{0}\psi -  \nu 
\psib \left( \sqrt{ 1 - \nu^{-2 } \del^2_1} -1 \right)  \psi
\,.  \label{dus}
\ea
 Indeed, the equation of motion for \rf{kus} written in 
 momentum space implies 
 $ p_0 =   { p^2_1  \ov  2 \nu + p_0 }$,  solved by  
 $p_0 = - \nu + \sqrt{ \nu^2 + p^2_1}$, which is the same 
  relation that follows from \rf{dus}.  The overall factor of $\nu$ 
  is finally absorbed by a redefinition of $\tau \to t= \nu \tau$, 
  i.e. we finish with \rf{rela}  where $J = \sqrt \l \ \nu$.

The Lagrangian  \rf{dus}
 is also equivalent to  the Dirac 
  Lagrangian  for a  massive  ($m = \nu = { J \ov
\sqrt \l}$)
relativistic 
2d fermion  with one component integrated out. 
This 
is not surprising  given that 
the  superstring  action in the 
$S^5$ geodesic (BMN)  background is known to contain free massive 
2d fermions \ci{metsa}. 
The usual 2d fermionic Lagrangian can be written as  
\be 
{\cal L} =  i \bar \Psi \rho^a \del_a \Psi  +  m \bar \Psi \rho^3 \Psi  \ , \ \ \ \ \ \ 
\bar \Psi= \Psi^\dagger \rho^0 \ , \ \ \ \ 
\rho^0=  i \sigma_2 \ , \ \ \ \ \ \rho^1=   \sigma_1 \ , \ee 
where  $\rho^3=\rho^0 \rho^1 = \s_3 $ and $\Psi= (\psi_1, \psi_2)$. Explicitly, 
 \be 
L= -  i  \psi^*_1  (\del_0 - \del_1)  \psi_1 -  i  \psi^*_2 ( \del_0  
+ \del_1)\psi_2
   + m ( \psi^*_1 \psi_2  + \psi^*_2 \psi_1) \ee
This leads   to the same    dispersion relation 
as the one that follows from  \rf{dus} (with only one solution chosen, 
as dictated by large mass expansion).  
 That means that  there should be a direct field redefinition that 
relates the two quadratic actions. 

As for possible higher order fermionic terms in \rf{dus}
(e.g. $\psib \psi \del_1 \psib \del_1 \psi$, etc.) 
we expect that there exists a field redefinition that 
completely eliminates them.  As suggested by the 
form of the exact solution of sect.
4.1,  such a field redefinition should  involve shifting  $\X_1$ 
by  fermionic terms.

\section*{Acknowledgments }

We are grateful to  G. Arutyunov, 
 N. Beisert, S. Frolov,  A. Fotopoulos,
 R. Metsaev, M. Kruczenski,  R. Roiban, M. Spardlin, 
 M. Staudacher and   A. Volovich
for useful discussions, suggestions  and comments.
 We would like also to thank the organizers and 
 participants of the 2004  ``QCD and String Theory''
KITP  workshop for a stimulating atmosphere, and 
   the KITP for the hospitality during  part of this work. 
The work of B.S. was supported by a Marie Curie Fellowship. 
The  work of A.T. was supported  by the DOE
grant DE-FG02-91ER40690, the INTAS contract 03-51-6346
and RS Wolfson award.  While visiting KITP this research was 
supported in part by the National Science Foundation under 
Grant No. PHY99-07949.


\setcounter{footnote}{0}
\setcounter{section}{0}
\appendix{The $\rho$-matrices}\label{appa}

We follow the notation of \ci{mt,mtt}.
The six $4 \times 4$  matrices   $\rho_{AB}^M$  are blocks of  
the $SO(6)$   Dirac   matrices $\gamma^M$
in the  chiral representation, i.e. 
\be\label{usgam}
\gamma^M
=\left(\begin{array}{cc}
 0   & (\rho^M)^{AB} 
 \\
 \rho_{AB}^M & 0
 \end{array}
 \right)\,,  \ \ \ \ \   \rho_{AB}^M =- \rho_{AB}^M\,,
 \qquad (\rho^M)^{AB}\equiv  - (\rho_{AB}^{M})^* \ , \ee
 \be 
 (\rho^M)^{AC}\rho_{CB}^N + (\rho^N)^{AC}\rho_{CB}^M
 =2\delta^{MN}\delta_B^A\, . 
 \ee
 Note that since  $X_M X^M =1$ 
 \be
X_{M}\rho^{M}_{AB}X_{N}\rho^{NBC}=\delta_{A}^{C}\,.
\ee
In this paper  we  have  chosen 
 the following representation for the $\rho^M_{AB} $ matrices
 \ba
\rho^1_{AB}&=&\left(\begin{matrix}0&0&0&1\\0&0&1&0\\0&-1&0&0\\-1&0&0&0\end{matrix}\right)\,,\ \ 
\rho^2_{AB}=\left(\begin{matrix}0&0&0&i\\0&0&-i&0\\0&i&0&0\\-i&0&0&0\end{matrix}\right)\,,\ \ \ 
\rho^3_{AB}=\left(\begin{matrix}0&0&-1&0\\0&0&0&1\\1&0&0&0\\0&-1&0&0\end{matrix}\right)\,,\nonumber \\
\rho^4_{AB}&=&\left(\begin{matrix}0&0&i&0\\0&0&0&i\\-i&0&0&0\\0&-i&0&0\end{matrix}\right)\,,\ \ 
\rho^5_{AB}=\left(\begin{matrix}0&1&0&0\\-1&0&0&0\\0&0&0&1\\0&0&-1&0\end{matrix}\right)\,,\ \ \
\rho^6_{AB}=\left(\begin{matrix}0&-i&0&0\\i&0&0&0\\0&0&0&i\\0&0&-i&0\end{matrix}\right)\,.\nonumber 
\ea
With this choice the following relations hold
\ba
X_M\rho^{M}_{ij}=\epsilon_{ijk}\X^{k}\,,\ \ \ \ \ 
X_M\rho^{M}_{i4}=\X_{i}\,,\ \ \ \ \ 
X_M\rho^{M}_{4j}=-\X_{j}\,,
\ea
where $i,\,j=1,2,3$ are the $SU(3)$ indicies, 
 $\epsilon_{123}=\epsilon^{123}=1$, and we have defined
\be
  \X_j =  X_{2j-1} + i X_{2j} \ , \ \ \ \ \ \ \ \ \ \ 
    \X^{i}\equiv \X_{i}^{*}\, , \ \ \ \ \ \  \X^i \X_i =1 \ . 
\ea
Similarly, we have
\be 
X_M\rho^{M}{}^{ij}=-\epsilon^{ijk}\X_{k}\,,\ \ \ \ \ \ 
X_M\rho^{M}{}^{i4}=-\X^{i}\,,\ \ \ \ \ \ \ \
X_M\rho^{M}{}^{4i}=\X^{i}\,.
\ee
We also define 
\be
\rho^{MN}{}^{A}{}_{B}=\ha\left(\rho^{M}{}^{AC}\rho^{N}{}_{CB}
-\rho^{N}{}^{AC}\rho^{M}{}_{CB}\right)\,.
\ee
With our choice of $\rho^M$   the only diagonal matrices among 
 $\rho^{MN}$ are $\rho^{12}$, $\rho^{34}$ and $\rho^{56}$.
  In the above $SU(3)$ notation we get 
 \be
 X_{M}\p X_{N}\rho^{MN}{}^{A}{}_{B}=\left(\begin{array}{cc}
 \X^{i}\p \X_{l}-\p \X^{i} \X_{l}+\delta_{l}^{i}\X_{m}\p \X^{m}&
  \epsilon^{ijk}\X_{j}\p \X_{k} \\
\epsilon_{ljk}\p \X^{j} \X^{k} & \X^{j}\p \X_{j}\end{array}
 \right)\,.
\ee
 Here  we have used that $\X_i \X^i=1$ implies 
 $\X^{j}\p \X_{j}=-\X_{j}\p \X^{j}$.
 
Other useful relations are 
(we always assume the sum over repeated $M,N$ indices) 
\ba
X^{M}(\rho^{Mi})^{l}{}_{k}&=2\delta^i_kX^l-\delta^l_kX^i\,,\qquad
X^{M}(\rho^{Mi})^{l}{}_{4}&=2\epsilon^{lmi}X_m\,,\qquad\nonumber \\
X^{M}(\rho^{Mi})^{4}{}_{k}&=0\,,\qquad\qquad\qquad\qquad\!\!\!\!
X^{M}(\rho^{Mi})^{4}{}_{4}&=X^i\,,\\
X^{M}(\rho^{M}{}_i)^{l}{}_{k}&=-2\delta^i_lX_k+\delta^l_kX_i\,,\qquad\!\!\!\!
X^{M}(\rho^{M}{}_i)^{l}{}_{4}&=0\,,\qquad\nonumber \\
X^{M}(\rho^{M}{}_i)^{4}{}_{k}&=2\epsilon_{kin}X^n\,,\qquad\qquad\qquad\!\!\!\!\!\!\!\!\!
X^{M}(\rho^{M}{}_i)^{4}{}_{4}&=-X_i\,.
\ea
Here we defined  
\be
\rho^{Mi}\equiv\rho^{M,2i-1}-i\rho^{M,2i}\,,\qquad\rho^{M}_i\equiv\rho^{M,2i-1}+
i\rho^{M,2i}\,.
\ee
We find also that 
\ba
X^M\eta_A(\rho^{Mi})^A{}_B\eta^B &=& 2\eta^i \eta_j\X^j+  
\X^i(\etab\eta  - \eta^j\eta_j ) -2\epsilon^{ijk}\X_k\eta_j\etab\,,\\
X^M\eta_A(\rho^{M}_{\ i}  )^A{}_B\eta^B&=& 2\eta_i \eta^j\X_j-   
\X_i(\etab\eta  - \eta^j\eta_j) -2\epsilon_{ijk}\X^k\eta^j\eta \,,
\ea
and
\ba
X^MX^K\rho^{MN}{}^i{}_j\eta_C\rho^{NK}{}^C{}_D\eta^D&=&
-2X_jX^i(\eta_k\eta^k-\eta^2) + 2\eta_jX^iX_k\eta^k - 2\eta^iX_j \eta_kX^k\nonumber \\&&
-2\epsilon_{kjm}X^i\eta X^m\eta^k
+ 2\epsilon^{k mi}X_mX_j\eta_k\etab \nonumber \\&&
+\delta^i_j(\eta_k\eta^k -\eta^2 - 2X^k\eta_kX_m\eta^m\,,\\
X^MX^K\rho^{MN}{}^i{}_4\eta_C\rho^{NK}{}^C{}_D\eta^D&=&
-2X_k\eta^k\epsilon^{kim}X_m\eta_k -2\eta\eta^i + 2\eta X^i X_k\eta^k\,,\\
X^MX^K\rho^{MN}{}^4{}_j\eta_C\rho^{NK}{}^C{}_D\eta^D&=&
2X^k\eta_k\epsilon^{jkm}X^m\eta_k -2\eta_j\etab + 2X_j X^k\eta_k\etab\,, \\
X^MX^K\rho^{MN}{}^4{}_4\eta_C\rho^{NK}{}^C{}_D\eta^D&=&
\eta^2-\eta_k\eta^k+2\eta_kX^k\eta^lX_l\,.
\ea
These formul\ae{} in turn give the relation used in  simplifying the quartic 
fermionic terms
in the string  action in  section~\ref{sec3} 
\ba
&&[X^M\eta_A(\rho^{MN})^A{}_B\eta^B][X^K\eta_C(\rho^{KN})^C{}_D\eta^D]
\nonumber \\   &=&
4\etab \eta \eta^i\eta_i  - 8\X_i\eta^i\X^j\eta_j\etab \eta 
+ 8\eta_i\X^i\eta^j\X_j\eta_k\eta^k-\eta_i\eta^i\eta_j\eta^j
 \nonumber \\
&+&  4\epsilon_{ijk}\eta^i\eta^j\X^k\eta_l\X^l\eta 
+ 4\epsilon^{ijk}\eta_i\eta_j\X_k\eta^l\X_l\etab\ . 
\ea

\setcounter{footnote}{0}
\setcounter{subsection}{0}
\appendix{ $SU(4)$ charges of  the string action }\label{appb}

Here  we  will express the string sigma model $SU(4)$ 
charges obtained in \ci{mt,mtt} in $SU(3)$ notation. 
The  $SU(4)$  charges 
are given by (using  our $AdS_5$ ansatz  \rf{adsansatz})
\ba
{\cal J}^A{}_B&=&\int d \s \  {\cal J}^0{}^A{}_B\,,
\\
{\cal J}^0{}^A{}_B&=&\frac{i}{2}X^M\p_0X^n\rho^{MN}{}^A{}_B\nonumber \\
&&
-\frac{\nu}{\sqrt{2}}\Bigl(\theta^A\theta_B+\eta^A\eta_B+\frac{1}{4}(\theta_C\theta^C+\eta_C\eta^C)
-\frac{1}{2}X^MX^K\rho^{MN}{}^A{}_B\eta_C\rho^{NK}{}^C{}_D\eta^D\Bigr)\,.\nonumber\\
\ea
Using the expressions in Appendix A 
 we can re-write  $ {\cal J}^0{}^A{}_B$   in the  $SU(3)$ notation
\ba
{\cal J}^0{}^i{}_j&=&\frac{i}{2}\Bigl(\X^i\p_0\X_j-\X_j\p_0\X^i+
\delta^i_j\X_k\p_0\X^k\Bigr)
\nonumber\\ &&
+\frac{\nu}{\sqrt{2}}\Bigl(\theta^i\theta_j+\eta^i\eta_j+
\X_j\X^i(\eta_k\eta^k+ \etab \eta )-2\eta_j\X^i\X_k\eta^k+2\eta^i\X_j\X^k\eta_k
\nonumber\\ &&\qquad\qquad\!\!\!\!
+2\epsilon_{kjm}\X^i\eta \X^m\eta^k-2\epsilon^{kmi}\X_j\eta_k \X_m\etab
\nonumber\\ &&\qquad\qquad\!\!\!\!
+\frac{1}{4}\delta^i_j(\theta_k\theta^k-  \thetab \theta-\eta_k\eta^k-
2\etab \eta +2\eta_k\X^k\X_m\eta^m)
\Bigr)
\,,\\
{\cal J}^0{}^i{}_4&=&\frac{i}{2}\epsilon^{imk}\X_m\p_0\X_k+\frac{\nu}{\sqrt{2}}
\Bigl(\theta^i\theta+2\eta^i\eta+\X^i\eta \X_k\eta^k-\epsilon^{imk}\X_k\eta_m \X_l\eta^l\Bigr)
\,,\\
{\cal J}^0{}^4{}_i&=&\frac{i}{2}\epsilon_{imk}\X^m\p_0\X^k+\frac{\nu}{\sqrt{2}}
\Bigl(\theta_i\thetab+2\etab\eta_i+\X_i\X^k\eta_k\etab-\epsilon_{imk}\X^k\eta^m \X^l\eta_l\Bigr)
\,,\\
{\cal J}^0{}^4{}_4&=&\frac{i}{2}\X^k\p_0\X_k+\frac{\n}{4\sqrt{2}}\Bigl(
3\thetab \theta +5\etab \eta +\theta_k\theta^k
+3\eta_k\eta^k-
\eta_k\X^k\eta^m\X_m\Bigr)\,,
\ea



\end{document}